%% file: main.tex
\documentclass[lettersize,journal]{IEEEtran}
\usepackage{amsmath,amsfonts}
\usepackage[linesnumbered,ruled,vlined]{algorithm2e}
\usepackage{pifont}
\usepackage{amsthm}
\usepackage{algpseudocode}
\usepackage{tcolorbox}
\usepackage{array}
\usepackage[caption=false,font=normalsize,labelfont=sf,textfont=sf]{subfig}
\usepackage{textcomp}
\usepackage{stfloats}
\usepackage{url}
\usepackage{verbatim}
\usepackage{graphicx}
\usepackage{ulem}
\usepackage{amsfonts}
\usepackage[utf8]{inputenc}
\usepackage{setspace}
\usepackage{cleveref}
\crefname{section}{§}{§§}
\Crefname{section}{§}{§§}
\usepackage{caption}
\usepackage{booktabs}
\usepackage{makecell}
\usepackage{MnSymbol}
\usepackage{utfsym}
\usepackage{filecontents, pgffor}
\usepackage{balance}
\usepackage{latexsym}
\usepackage{color,xcolor,colortbl}
\usepackage{multirow}

\definecolor{mySourceNodecolor}{HTML}{ddf1e8}
\definecolor{myOtherNodecolor}{HTML}{385775}
 \usepackage{tikz}

\usepackage{listings, xcolor}
\usepackage{amsmath}

\definecolor{verylightgray}{rgb}{.97,.97,.97}

\lstdefinelanguage{Solidity}{
	keywords=[1]{anonymous, assembly, assert, balance, break, call, callcode, case, catch, class, constant, continue, constructor, contract, debugger, default, delegatecall, delete, do, else, emit, event, experimental, export, external, false, finally, for, function, gas, if, implements, import, in, indexed, instanceof, interface, internal, is, length, library, log0, log1, log2, log3, log4, memory, modifier, new, payable, pragma, private, protected, public, pure, push, require, return, returns, revert, selfdestruct, send, solidity, storage, struct, suicide, super, switch, then, this, throw, transfer, true, try, typeof, using, value, view, while, with, addmod, ecrecover, keccak256, mulmod, ripemd160, sha256, sha3}, % generic keywords including crypto operations
	keywordstyle=[1]\color{blue}\bfseries,
	keywords=[2]{address, bool, byte, bytes, bytes1, bytes2, bytes3, bytes4, bytes5, bytes6, bytes7, bytes8, bytes9, bytes10, bytes11, bytes12, bytes13, bytes14, bytes15, bytes16, bytes17, bytes18, bytes19, bytes20, bytes21, bytes22, bytes23, bytes24, bytes25, bytes26, bytes27, bytes28, bytes29, bytes30, bytes31, bytes32, enum, int, int8, int16, int24, int32, int40, int48, int56, int64, int72, int80, int88, int96, int104, int112, int120, int128, int136, int144, int152, int160, int168, int176, int184, int192, int200, int208, int216, int224, int232, int240, int248, int256, mapping, string, uint, uint8, uint16, uint24, uint32, uint40, uint48, uint56, uint64, uint72, uint80, uint88, uint96, uint104, uint112, uint120, uint128, uint136, uint144, uint152, uint160, uint168, uint176, uint184, uint192, uint200, uint208, uint216, uint224, uint232, uint240, uint248, uint256, var, void, ether, finney, szabo, wei, days, hours, minutes, seconds, weeks, years},	% types; money and time units
	keywordstyle=[2]\color{teal}\bfseries,
	keywords=[3]{block, blockhash, coinbase, difficulty, gaslimit, number, timestamp, msg, data, gas, sender, sig, value, now, tx, gasprice, origin},	% environment variables
	keywordstyle=[3]\color{violet}\bfseries,
	identifierstyle=\color{black},
	sensitive=true,
	comment=[l]{//},
	morecomment=[s]{/*}{*/},
	commentstyle=\color{gray}\ttfamily,
	stringstyle=\color{red}\ttfamily,
	morestring=[b]',
	morestring=[b]"
}

\begin{document}
\title{Penetrating the Hostile: Detecting DeFi Protocol Exploits through Cross-Contract Analysis}

\author{Xiaoqi Li, Wenkai Li, Zhiquan Liu, Yuqing Zhang, Yingjie Mao
\thanks{Xiaoqi Li, Wenkai Li, Yingjie Mao are with the Hainan University, Haikou, 570228, China (e-mail: csxqli@ieee.org, cswkli@hainanu.edu.cn, yingjiemao@hainanu.edu.cn)}
\thanks{Zhiquan Liu is with the Jinan University, Guangzhou, 510632, China (e-mail: zqliu@jnu.edu.cn)}
\thanks{Yuqing Zhang is with the University of Chinese Academy of Sciences, Beijing, 100049, China (e-mail: zhangyq@nipc.org.cn)}
\thanks{Wenkai Li (cswkli@hainanu.edu.cn) is the corresponding author.}
\thanks{This manuscript is an extended version of our work~\cite{li2024defitail}. It has been extended more than 40\% over the WWW conference version, including: (1) Optimization of the background and motivation of the framework (Sec. II)
(2) Elaboration on the detailed principles of the framework (Sec. III).
(3) Enhancement of the analysis of the experiments (Sec. IV).
(4) Addition of discussion with the exploits detected by our framework (Sec. V).
 }
}

\markboth{IEEE Transactions on Information Forensics and Security,~Vol.~XX, No.~X, XXX~2025}%
{Li \MakeLowercase{\textit{et al.}}: Detecting DeFi Protocol Exploits through Cross-Contract Analysis}
\newcommand{\blackding}[1]{\ding{\numexpr181+#1\relax}}
\newcommand{\whiteding}[1]{\ding{\numexpr171+#1\relax}}

\maketitle

\begin{abstract}
Decentralized finance (DeFi) protocols are crypto projects developed on the blockchain to manage digital assets. Attacks on DeFi have been frequent and have resulted in losses exceeding \$80 billion. Current tools detect and locate possible vulnerabilities in contracts by analyzing the state changes that may occur during malicious events. However, this victim-only approaches seldom possess the capability to cover the attacker's interaction intention logic. Furthermore, only a minuscule percentage of DeFi protocols experience attacks in real-world scenarios, which poses a significant challenge for these detection tools to demonstrate practical effectiveness. In this paper, we propose DeFiTail, the \textit{first} framework that utilizes deep learning technology for access control and flash loan exploit detection. Through feeding the cross-contract static data flow, DeFiTail automatically learns the attack logic in real-world malicious events that occur on DeFi protocols, capturing the threat patterns between attacker and victim contracts. Since the DeFi protocol events involve interactions with multi-account transactions, the execution path with external and internal transactions requires to be unified. Moreover, to mitigate the impact of mistakes in Control Flow Graph (CFG) connections, DeFiTail validates the data path by employing the symbolic execution stack. Furthermore, we feed the data paths through our model to achieve the inspection of DeFi protocols. Comparative experiment results indicate that DeFiTail achieves the highest accuracy, with 98.39\% in access control and 97.43\% in flash loan exploits. DeFiTail also demonstrates an enhanced capability to detect malicious contracts, identifying 86.67\% accuracy from the CVE dataset. By monitoring existing contracts, we identified five distinct categories of vulnerabilities: repetition abuse, unsafe unintended exploitation, signature violated exploitation, insecure interfaces exploitation, and unrestricted token transfer.
\end{abstract}

\input{sections/intro}

\input{sections/background}

\input{sections/method}
\input{sections/evaluation}

\input{sections/relatedwork}

\input{sections/conclusion}

\section{ACKNOWLEDGMENTS}
This work is sponsored by the National Natural Science Foundation of China (No.62362021 and No.62402146), CCF-Tencent Rhino-Bird Open Research Fund (No.RAGR20230115), and Hainan Provincial Department of Education Project (No.HNJG2023-10).

\balance

\bibliographystyle{IEEEtran}
\normalem
\bibliography{main}

\vspace{-3em}
\begin{IEEEbiography}[{\includegraphics[width=1in,height=1.25in,clip,keepaspectratio]{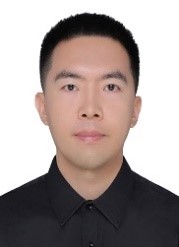}}]{Xiaoqi Li}
is an associate professor at Hainan University. Previously, he was a researcher at the Hong Kong Polytechnic University. He received his Ph.D. in Computer Science from Hong Kong Polytechnic University, MSc in Information Security from the Chinese Academy of Sciences, and BSc in Information Security from Central South University. His current research interests include Blockchain/Mobile/System Security and Privacy, Ethereum/Smart Contract, Software Engineering, and Static/Dynamic Program Analysis. He received best paper awards from INFOCOM’18, ISPEC’17, CCF’18, and an outstanding reviewer award from FGCS’17.
\end{IEEEbiography}
\vspace{-3em}
\begin{IEEEbiography}[{\includegraphics[width=1in,height=1.25in,clip,keepaspectratio]{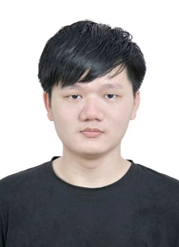}}]{Wenkai Li} is currently pursuing a doctor’s degree in the School of Cyberspace Security at Hainan University, China. Previously, he received a master's degree in the School of Cyberspace Security at Hainan University. His research lies in smart contract security and malicious behavior analysis, focusing on enhancing blockchain security through software and data analytics. He is also exploring the integration of artificial intelligence, such as graph neural networks and large language models.
\end{IEEEbiography}
\vspace{-3em}
\begin{IEEEbiography}[{\includegraphics[width=1in,height=1.25in,clip,keepaspectratio]{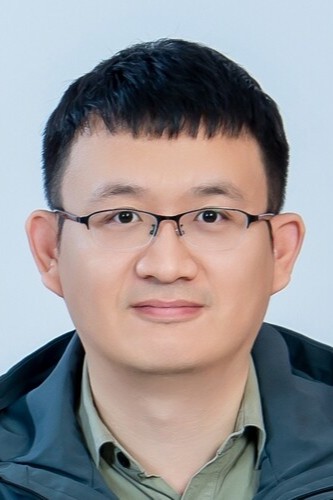}}]{Zhiquan Liu} received the B.S. degree from the
School of Science, Xidian University, Xi’an, China, in 2012, and the Ph.D. degree from the School of Computer Science and Technology, Xidian University, in 2017. He is currently a Full Professor, the Doctoral Supervisor, and the Deputy Dean of the College of Cyber Security, Jinan University, Guangzhou, China. His current research focuses on security, trust, privacy, and intelligence in vehicular networks. He currently serves as the area editor, associate editor, or academic editor of more than 10 SCI-index journals, such as IEEE Transactions on Industrial Informatics, IEEE Internet of Things Journal, Information Fusion, IEEE Network, etc.
\end{IEEEbiography}
\vspace{-3em}
\begin{IEEEbiography}[{\includegraphics[width=1in,height=1.25in,clip,keepaspectratio]{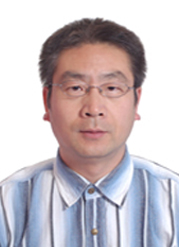}}]{Yuqing Zhang}
is the Director of the Chinese National Computer Network Intrusion Prevention Center, Deputy Director of the Chinese National Engineering Laboratory of Computer Virus Prevention Technology, Vice Dean of the School of Computer and Control Engineering at the Chinese Academy of Sciences, and Professor at Hainan University. He received his Ph.D. from Xi'an University of Electronic Science and Technology. He has presented over 100 papers and 7 national/industry standards. His current research interests include Network Attacks and Prevention, Security Vulnerability Mining and Exploitation, IoT System Security, AI Security, Data Security, and Privacy Protection. 
\end{IEEEbiography}
\vspace{-3em}
\begin{IEEEbiography}[{\includegraphics[width=1in,height=1.25in,clip,keepaspectratio]{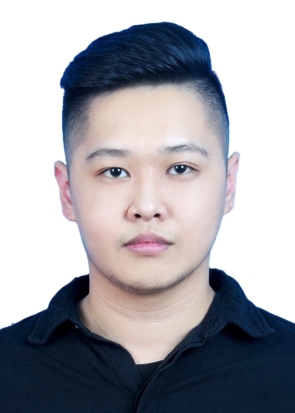}}]{Yingjie Mao}
is currently pursuing a master’s degree in the School of Cyberspace Security at Hainan University, China. Previously, he received a B.E. degree from the Southwest University of Science and Technology. His current research interests include Blockchain Security/Privacy and Large Language Models.
\end{IEEEbiography}

\vfill

\end{document}

%% file: sections/intro.tex
\section{Introduction}
\label{sec:intro}

Recently, Ethereum-compatible blockchains have witnessed a significant surge in popularity~\cite{Etherscan}. This increase can be primarily attributed to the growth of ecosystems, such as Decentralized Finance (DeFi)~\cite{werner2022sok,zhou2023sok} and Non-Fungible Tokens (NFTs)~\cite{das2022understanding}, established on these blockchains. These ecosystems, comprising Decentralized Applications (DApps)~\cite{singh2023dapps} that execute diverse functions via smart contracts, have garnered a substantial number of active users and market assets. However, security vulnerabilities could exist in any software~\cite{aggarwal2005software,li2025interaction,li2025sweeping}. As the logical component of DApp, the vulnerability of smart contracts directly affects the security of massive digital assets on users. According to statistics, DeFi-related security incidents have accounted for losses exceeding \$80 billion. While access control remains a common vulnerability, the techniques employed to execute such attacks have evolved to become increasingly sophisticated. For example, the attacker found that the verification of the withdraw function within the Orbit Chain contract was inadequate~\cite{DigiFinex2024Orbit}. Through exploiting fake signatures to satisfy the verification threshold, the attacker managed to steal multiple tokens, such as 9,500 ETH and 231 wBTC (with a total value of more than \$81.5 million).

Previous research \cite{luu2016making,chen2018detecting,he2019learning} focused on detecting vulnerabilities in smart contracts, utilizing user-defined rules and expert knowledge to standardize detection capabilities. Moreover, the success of deep learning models builds different embeddings, which has been proven in identifying correlation from historical contracts~\cite{chen2020finding,gao2019smartembed}, enabling them to discover vulnerable patterns that can capture fragile contracts.

Semantic embedding, which constructs different forms of representation vectors through semantic rules, has been utilized to detect many vulnerabilities. The semantic rules refer to the special rules defined according to the vulnerability or the characteristics of the code, such as the integration of code into control flow based on syntax rules~\cite{liu2021combining}, the conversion of code into a tree structure according to call flows~\cite{wang2024contractcheck}, and the incorporation of contract transfer into a graph structure corresponding to token flows~\cite{yuan2023enhancing,wu2023defiranger}. These rules construct various vector features by analyzing contract semantics. The approach of semantic embedding has been proven to be helpful for detecting multiple vulnerabilities, such as integer overflow, timestamp dependency, reentrancy, DoS vulnerability, and block state dependency. Initially, embedding processes involved the direct conversion of source code or opcode into distinct representations, followed by using deep learning models to learn sequential textual features~\cite{gao2019smartembed,gao2020deep,li2022poster}. Subsequently, more advanced techniques leverage structured semantic information, such as program paths extracted from Control Flow Graphs (CFG), to enhance detection performance. The graph neural networks~\cite{liu2021combining,zhuang2021smart} or language models~\cite{SmarTest2021UsenixSunbeom,wu2021peculiar} are utilized, enabling the classification of code snippets for vulnerability presence. However, it is insufficient for the vulnerability detection of smart contracts. These methods are unaware of capturing accurate attack processes, particularly when multiple contracts are involved, and are incapable of detecting potential vulnerabilities involving multiple contracts. Therefore, there is potential for improvement in the program path representation offered by a single contract when dealing with scenarios that involve multiple contracts. 

\begin{figure}[!t]
\centering
\includegraphics[width=\linewidth]{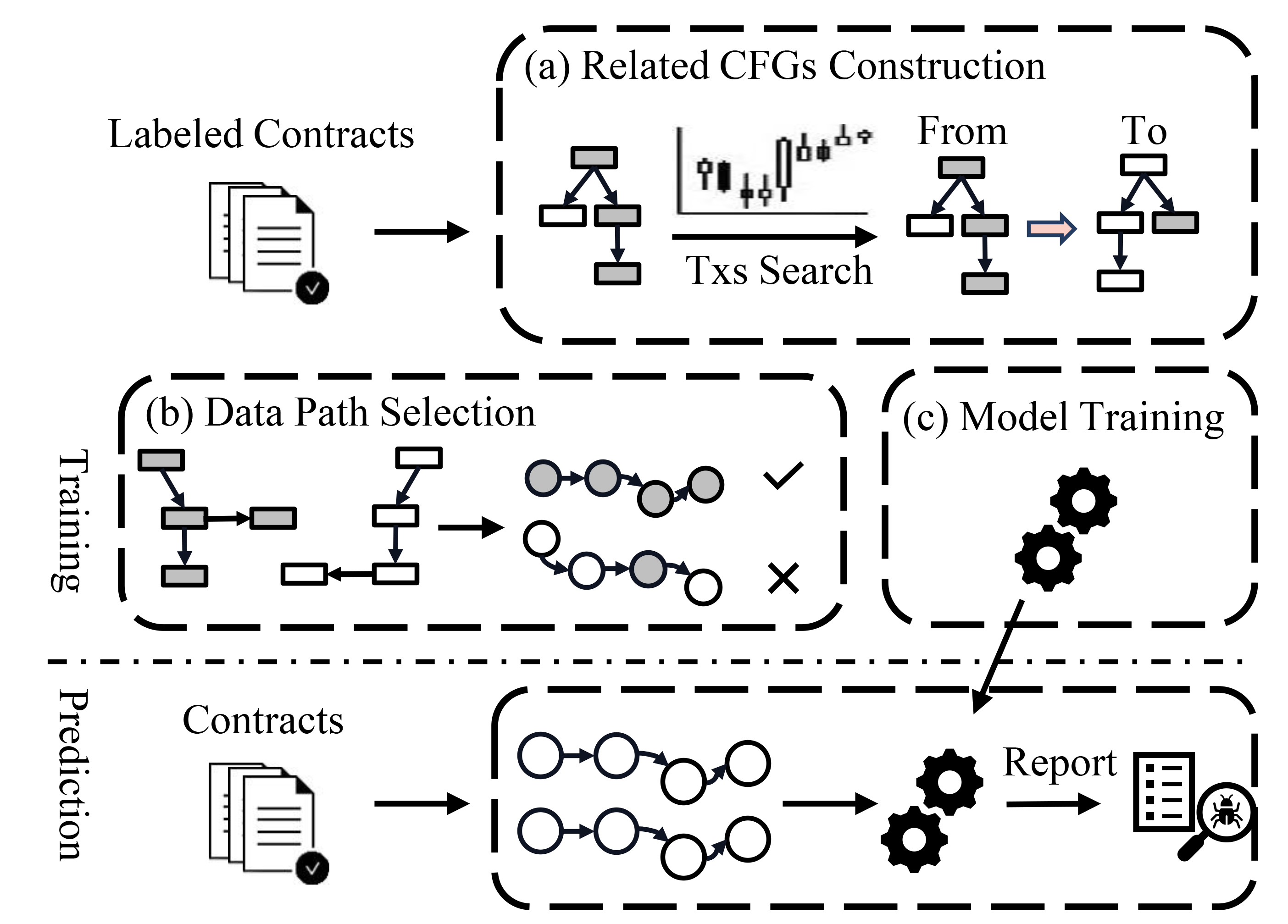}
\caption{Overview of DeFiTail. Above the dotted line is the training phase, and below the dotted line is the vulnerability prediction stage.}
\label{fig:framework}
\vspace{-2ex}
\end{figure}

Deep learning has demonstrated superior performance compared to traditional methods across a wide range of fields~\cite{kong2025uechecker,zhang2025penetration,gong2025information}. Data-driven could be effectively employed to leverage these advancements, feeding data into a model to learn the implicit features in contracts.
However, there are still some challenges in analyzing DeFi projects. 

\noindent$\bullet $\textbf{ Challenge 1 (C1): Invocation Pattern Learning.} Most execution processes in malicious DeFi events involve invocations of multiple contracts~\cite{REKT-database2023solidity}. However, the conventional approaches~\cite{gao2019smartembed,gao2020deep,zhuang2021smart} focus on examining isolated data paths or execution paths within individual contracts, which is insufficient for inspecting DeFi protocols that contain lots of invocation patterns in multiple contracts.  

\noindent$\bullet $\textbf{ Challenge 2 (C2): External and Internal Path Unification.} Data paths in DeFi protocols involve transaction-level external paths and code-level internal paths~\cite{zhang2020txspector}, where the external path means the execution flow between contracts, and the internal path means the control flow in a single contract. However, current studies~\cite{zhang2020txspector,wang2021blockeye, SmarTest2021UsenixSunbeom,ramezany2023midnight} focus on unifying contracts at the transaction level, lacking unifying external paths of DeFi protocols at the code level.

\noindent$\bullet $\textbf{ Challenge 3 (C3): Data Path Feasibility Validation.} Previous studies~\cite{chen2019large,ma2021pluto} have revealed limitations in comprehending the data paths extracted from the CFG. The smart contracts in a DeFi project can generate several CFGs, yielding multiple data paths after the CFGs connection. However, the different choices of entry points can determine different data paths extracted from the connected CFGs, generating different operational sequences. Therefore, it is essential to validate the feasibility of data paths at the operation level.

\textbf{Our Solution.} 
To address these challenges, we implement DeFiTail, which is a specialized DeFi inspection framework that learns the interaction patterns in data paths with deep learning. As shown in Figure \ref{fig:framework}, DeFiTail consists of three parts corresponding to the solutions below. Parts (a), (b), and (c) represent the modules S2, S3, and S1 respectively.

\noindent$\bullet $\textbf{ Solution to C1 (S1)}: We take advantage of sequence and graph learning technology, extracting sequential execution process features and structural heterogeneous graph features. First, we convert the data paths that contain opcodes and operands to a series of sequences according to the execution order in the connected CFGs. Then, a sequence learning model learns the sequence features, and a graph learning model is used to build the heterogeneous graph for extracting the structural features. By combining the sequence and structural features, a more complete feature can be extracted.

\noindent$\bullet $\textbf{ Solution to C2 (S2)}: We analyze the external transactions in the Ethereum Virtual Machine (EVM), paralleling the internal transaction logic in the smart contract. Through the function segmentation with CFG construction, the 4-byte function signature in the caller function is obtained. Then, we assess whether an invocation exists between functions by examining the presence of the function signatures within another function, unifying external paths and internal paths.
 
\noindent$\bullet $\textbf{ Solution to C3 (S3)}: We integrate a symbolic execution stack into DeFiTail, enabling the validation of data path feasibility. The symbolic execution stack uses symbols to record the number of bytes in the stack. Comparing the equal relation between stack height and opcode rules in EVM can determine whether the path is feasible.

The main contributions of this paper are as follows:
\begin{itemize}
    \item  To the best of our knowledge, we propose the first inspection framework DeFiTail, to detect DeFi attacks from the perspective of interaction between attacker and victim. DeFiTail can detect access control and flash loan exploits on various EVM-compatible blockchains (\cref{sec::method}).
    \item We unify external and internal paths, and connect CFGs between bytecode contracts. Furthermore, the data path validation is formulated into the path reachability problem, identifying feasible data paths (\cref{subsec::dps}).
    \item We evaluate DeFiTail and it outperforms SOTAs by 16.57\% and 11.26\% points in detecting access control and flash loan exploits. Moreover, we explore the performance enhancements between CFG connection and data path validation (\cref{sec::evaluation}). 
    \item We open-source the data and codes of DeFiTail at \url{http://doi.org/10.6084/m9.figshare.24117993}. 
\end{itemize}

% https://figshare.com/s/3e2eca154a1a66c35225

%% file: sections/background.tex
\section{Background and Motivation}
\label{sec:background}

In this section, we discuss the background of the contract transaction and exploitation definition. We also provide a motivating example.

\subsection{Contract Transaction}
Transactions are issued by accounts, Externally Owned Accounts (EOA) and contract accounts, to update the state of the blockchain, e.g., token transfers between two users. Blockchain refers to Ethereum in this paper, one of the most popular blockchain systems. The operation of reversing the state of the block (e.g., growing balance) is stored in the form of transactions in all blockchain nodes. In EVM, the contract does not directly initiate transactions. EOA account initiates an external transaction, triggering the execution of the logic in the contract. Then the function calls within the contract result in cross-contract interactions (i.e., internal transactions).

An internal transaction records a relationship that occurs within a contract, including calling another contract and creating a new contract. Based on the internal transactions, the EVM converts the logic source contract into deployed instructions, and then performs operations to modify the state of the contract. Through the variability of the state, the correct execution logic in contracts can be guaranteed.

External transaction refers to the action initiated by an EOA. To explore the transaction patterns related to DeFi protocol attacks, we analyze the transactions that interact with the deployed contracts, i.e., the recipient is the contract. Since the contract cannot send transactions, and the contract creation only contains the bytecode of contracts, we focus only on the transactions with contracts. 

When a contract interacts with another one by a function call operation, the data flows or call flows are formed. As shown in the Table \ref{tab:terms_definition}, Data flow means that the data execution path follows the sequential execution order. Call flow means the path in a call connection order between two contracts or CFGs. Function calls between two contracts require verification of the function signature through the function selector. Since we focus on EVM-level contracts to ensure adaptability, we analyze the assembly of operation instructions from bytecode. Instructions or opcodes, such as \textsc{delegatecall} and \textsc{callcode}, access the calling contract's storage and check the called contract's signature.

\begin{table}[ht]
% \vspace{-1ex}
\centering
\small
\caption{The Terms Definition}
\vspace{-1ex}
\label{tab:terms_definition}
\begin{tabular}{l | p{6cm}}
    \hline
   \textbf{Terms} &  \textbf{Description} \\ 
    \hline
    Path          &  The sequence of opcodes in a specific order. \\
    Data Flow     &  The sequence of opcodes by the execution order of contracts.\\
    Call Flow     &  The path in a call connection order between two contracts or CFGs.  \\
    Control Flow  &  The path with the contract execution in a CFG. \\ 
    DPS Path      &  The path with the DPS order of the CFG or connected CFG(rCFG).\\ 
    \hline 
\end{tabular}
\vspace{-2ex}
\end{table}

\subsection{Exploitation Definition}
\noindent\textbf{Access Control.} Access control refers to programs that fail to effectively manage permission assignments, allowing unauthorized users to access sensitive data or execute critical operations. Attacker $A$ achieves the transaction initiation $AT$, sending the transactions. Then victim $B$ receives and executes the transaction (i.e., $Tran(B)$). So the Attacker $A$
may exploit vulnerabilities $Defects(\delta[B]_c)$ inherent in the contract code $\delta[B]_c$ of victim $B$, such as poorly designed role permissions, insufficient checks for functional restrictions, misuse of security libraries, reliance on insecure functions, and inadequate oversight of contract escalation. Then, attacker $A$ can achieve the data access behavior $Access(A, \delta[B]_d)$, and access the sensitive data $\delta[B]_d$ of victim $B$.
The rule of access control exploitation is shown as follows,
\[
\frac{
\begin{array}{c}
AT, Tran(B), Defects(\delta[B]_c), Access(A, \delta[B]_d)
\end{array}
}{
Access\ Control\ Exploitation
}
\] 
These access control defects can lead to severe consequences, including financial losses, unauthorized disclosure of sensitive information, and unauthorized modifications to contract status.

\noindent\textbf{Flash Loan Exploitation.} Flash loan exploitation refers to the use of the flash loan mechanism for malicious purposes. Flash loans are collateral-free loans that enable users to borrow substantial amounts of money within a single transaction, with the requirement to repay the loan before the transaction is finalized. Attacker $A$ can leverage a flash loan to quickly acquire large sums of money, exploiting malicious logic $\delta[A]_c$, such as contract defects and market manipulation, to influence market tokens and price fluctuations, ultimately generating profits after the repayment of the flash loan. 
The rule of the flash loan attack is shown as follows.
\[
\frac{
\begin{array}{c}
AT,  FlashLoan(T), MaliciousLogic(\delta[A]_c), \\
Repayment(T)
\end{array}
}{
Flash\ Loan\ Exploitation
}
\]

The key feature of flash loan exploitation is that both borrowing and repayment must occur in a unified transaction.

\subsection{Motivating Example}
\label{subsec:motivating_example}
%%  攻击示例

\begin{figure}[ht]
\centering
\includegraphics[width=\linewidth]{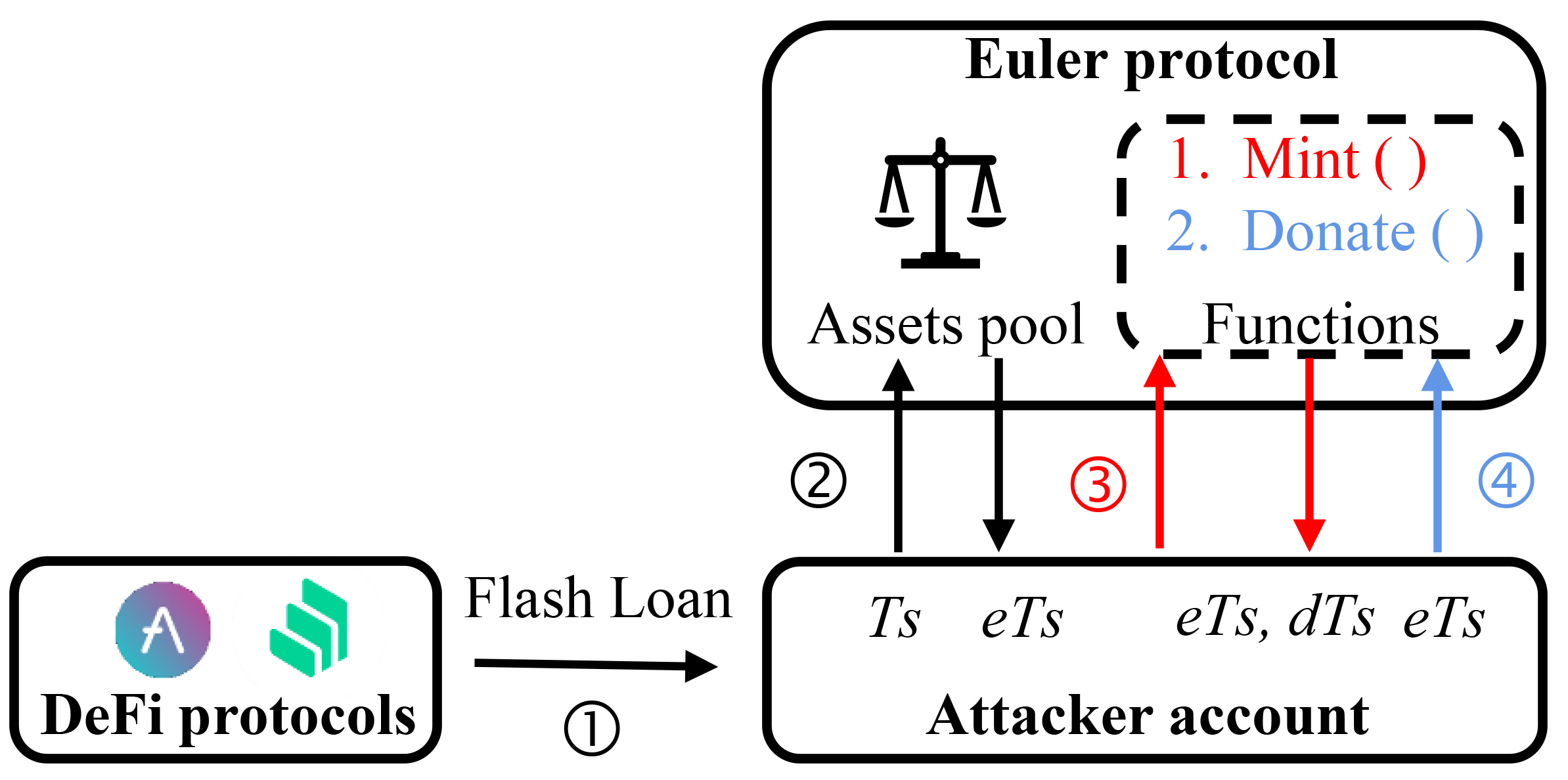}
\caption{An Motivating Example of Attacks Process in Euler Protocol. \ding{172}-\ding{175} steps lead to insolvency of the attacker.}
\label{fig:attack_flows}
\end{figure}

Figure \ref{fig:attack_flows} depicts the assault on the Euler protocol, which resulted in a loss of \$195 million on March 13, 2023. Accounts include EOAs and contract accounts, and to better describe the overall logic, the attacker contract is represented by the account. Flash loan is a DeFi lending mechanism that permits users with fewer assets and deposits to be collateralized, and the procedure is carried out through smart contracts. In Figure \ref{fig:attack_flows}, the attacker's contract account borrowed a certain amount of tokens (\textit{Ts}) through a flash loan, and through a series of token conversions with the Euler protocol, it eventually led to bad debt (step \ding{173}-\ding{175}). Among them, \textit{eTs} stands for collateral tokens, and \textit{dTs} stands for debt tokens. In step \ding{173}, the attacker deposits \textit{Ts} into the Euler protocol and exchanges a certain amount of \textit{eTs}. Then, by calling the \texttt{Mint()} function in step \ding{174}, more \textit{eTs} and \textit{dTs} with corresponding proportions are obtained to increase asset liquidity. Finally, the attacker calls the \texttt{Donate()} function in step \ding{175} and pays a certain amount of \textit{eTs} so that the state of \textit{eTs} below \textit{dTs} reaches the liquidation condition. Furthermore, as \textit{eTs} exceed \textit{dTs}, an additional amount remains after repaying the debt. These extra \textit{eTs} seize more collateral than initially intended to repay the flash loan, resulting in a profit for the attacker and a loss in Euler protocol. Therefore, the attack mainly stemmed from failing to check whether the assets were secure in step \ding{174}, resulting in the liquidation of the assets after the donation. The rule of the above exploitation is shown as follows,
\[
\frac{
\begin{array}{c}
FlashLoan(T), \exists f \in Func(C), TokenSwap(f),\\
 ManipulateState(f), \neg ValidateCollateral(f)\\
\end{array}
}{
Flash\ Loan\ Exploitation\ in\ Euler\ Finance
}
\]

However, the above is only a special attack on the Euler protocol. To correctly learn the specific expression of the threat pattern, and apply the pattern to automatically detect DeFi protocols, our fundamental idea is to leverage the benefits of program analysis and deep learning technologies. We first construct the control flow graph of various contracts, and collect the data flow through program analysis technology. Subsequently, we use deep learning technology to learn threat patterns from the data flows. When analyzing the attack process from the perspective of semantic embedding, it is necessary to focus on the control flows between different contracts, representing the execution logic within the contracts. Due to the fact that the attacks are a dynamic process involving state transitions between multiple contracts. Therefore, we need to deal with the bytecodes in the possible data flow of different contracts through program analysis technology. Furthermore, the patterns of attacks can be learned through deep learning technology, detecting possible attacks on contracts of DeFi protocols. Summarizing these motivations, the program analysis and AI become indispensable in our framework.

%% file: sections/method.tex
\section{The Proposed Framework}
\label{sec::method}

In this section, we introduce our method, which comprises three key components. First, \cref{subsec::rcc} transforms the relevant contracts into CFGs and connects them. Next, \cref{subsec::dps} extracts the feasible data path by tracking the execution logic of the connected CFG. Finally, \cref{subsec::model} leverages advanced deep learning algorithms to learn and detect the implicit patterns in the data path. Specifically, the CFG is an abstract representation of program execution flow. The data flow is the path through which data flows in a CFG, and we will use the data
flow path interchangeably. We divide code into blocks without call relationships, where the block is called a segment and the flow of calls between segments is called call flow. The control flow refers to a path (i.e., the sequence of opcodes during the execution of contracts) in a CFG.

\subsection{Related CFGs Construction}
\label{subsec::rcc}

When establishing an internal CFG $G=(N, E)$ within a single contract $CA$, the initial step involves partitioning basic blocks based on jumping or stopping operations (i.e., \texttt{STOP}, \texttt{SELFDESTRUCT}, \texttt{RETURN}, \texttt{REVERT}, \texttt{INVALID}, \texttt{SUICIDE}, \texttt{JUMP}, and \texttt{JUMPI}). Note that the basic blocks are represented as the nodes $N \in G$, and the jumping directions mean the edges $E \in G$. Each node $n \in N$ can be represented as a triple-group $n=(id, type, code)$, where the $id$ is the identification, $type$ indicates the type of the node (i.e., starting, ending, and conditional node), $code$ stores the opcodes or operations in the basic block. Subsequently, data flow is orchestrated in alignment with jumping operations (i.e., \texttt{JUMP} and \texttt{JUMPI}). Specifically, the calldata in the jumping operations contain the function signatures, enabling coherent division by discerning function entries based on varying signatures.
 
However, constructing a CFG for a single contract faces a challenge. It lacks division of the basic block by calling instructions, disregarding potential interactions involving the invocation of other contracts. The CFG effectively concatenates the sequential execution of all inter-functions. While the external jumping branches exist, the inter-CFG flow can not be completely covered. Specifically, the CFG constructed by single-contract inadequately captures the logic in the multi-contract call scenarios.

Function signatures can be either explicit or implicit. In the case of explicit function signatures, the function signature is stored in the first 4 bytes of the calldata data. This means that we can easily compare whether the hash ID exists in the contract. However, the implicit function signature cannot be ascertained directly, owing to the possibility of the signature changing due to certain operation rules.

\begin{figure}[t]
\centering
\includegraphics[width=\linewidth]{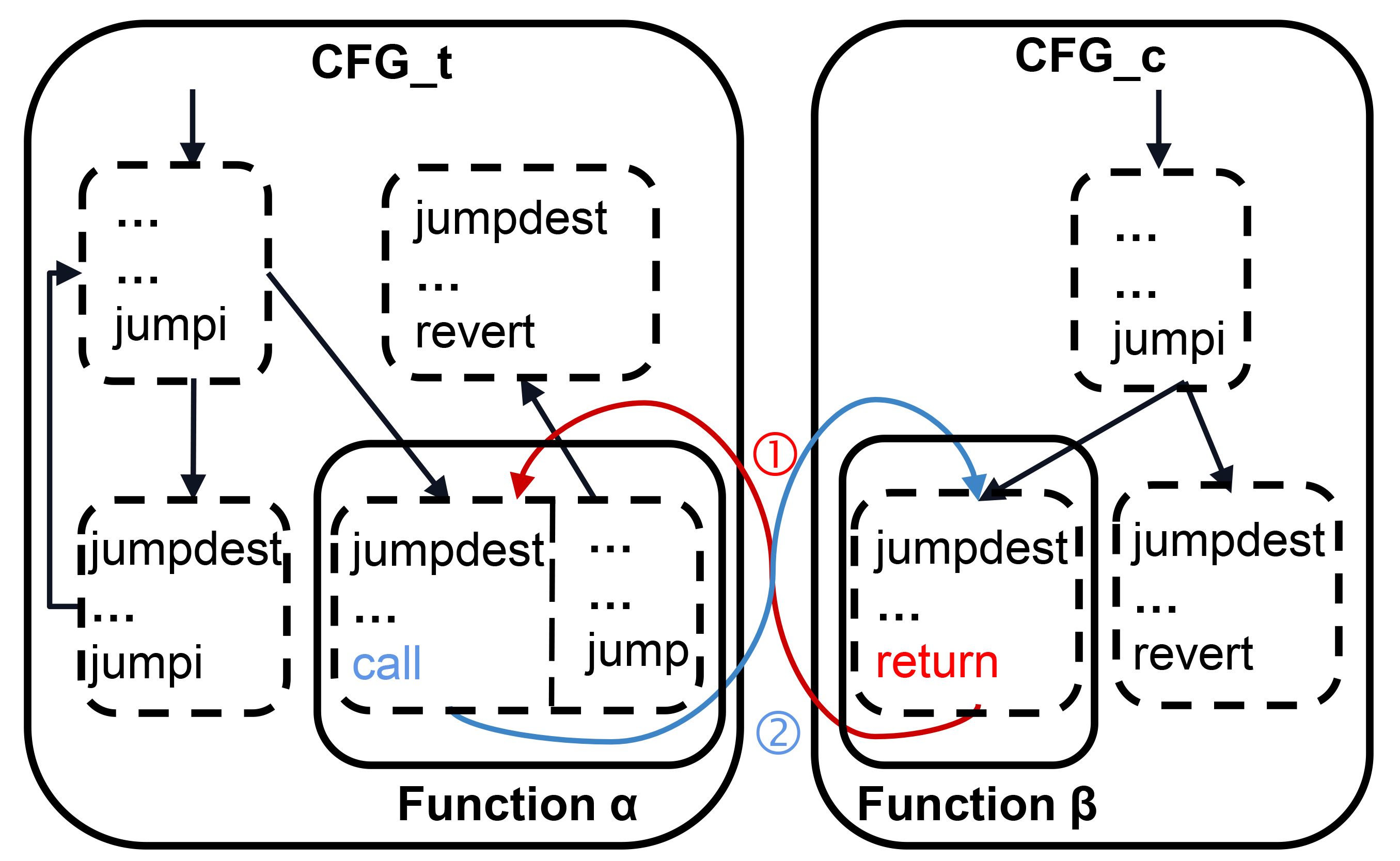}
\caption{A Motivating Example of Related CFG Construction. \ding{172} and \ding{173} represent the sequence of contracts connection.}
\label{fig:related_CFG_construction}
\vspace{-4ex}
\end{figure}

Therefore, inspired by the construction of the single-contract CFG~\cite{kong2023defitainter,ma2021pluto}, we slice the contract snippets with jumping and stopping operations. The calling operations (i.e., \texttt{CALL}, \texttt{DELEGATECALL}, \texttt{STATICCALL}, \texttt{CALLCODE}) and the return operations (i.e., \texttt{RETURN}) are leveraged as foundational flags to delineate the control flows between contracts. As depicted in Figure \ref{fig:related_CFG_construction}, we categorize relevant contracts based on their functions and meticulously capture the function signature data for each constituent function throughout the process. When a function involves a calling operation, we divide the function into two segments. The segments are guided by the corresponding instruments (e.g., \texttt{CALL}, \texttt{DELEGATECALL}, etc.). In the context of the \texttt{CALLDATA}, a function signature is located within the call data to which the calling operation is transitioned. In the instance presented in Figure \ref{fig:related_CFG_construction}, the calling operations are recognized for function $\alpha$ within $CFG_t$. While the function interface of the function $\beta$ within $CFG_c$ matches the signature incorporated in the calldata of function $\alpha$. The execution of steps \ding{172} and \ding{173} then proceeds to insert a control flow graph node, connecting different CFGs.

To address these challenges, we have implemented several vital discrepancies. As shown in Algorithm \ref{algo:getRCFG}, we have implemented a function-based method for contract slicing, departing from the traditional basic block segmentation. Moreover, we identify functions culminating with a \texttt{RETURN} opcode within the callee contract $\mathbb{B}$, while slicing functions containing the calling operation in the caller function $\mathbb{A}$. Furthermore, the $\perp$ indicates that the function path is feasible, $W$ represents the dictionary-type entry block nodes in CFG, and $O$ means the dictionary that stores the following blocks. Within the instance of $\mathbb{A}$, an inquisition is conducted to establish the presence of the function signatures within $\mathbb{B}$, thereby ensuring the occurrence of external calls. This approach offers the advantage of initiating the analysis directly from the function signature of $\mathbb{B}$, thus eliminating exhaustive internal data overhead.

Specifically, we investigate the occurrence of a designated function signature within a specified bytecode contract. Notably, the function signatures are uniquely represented by 4-byte hash IDs. However, a previous study~\cite{chen2019large} shows that the intricacies inherent exist in the bytecode-to-opcode conversion process (e.g., swap and shift), causing the direct alignment of these function signatures between the bytecode and opcode of the contract is impeded. To address this challenge, we propose a methodology that involves the conversion of bytecodes into opcodes within the context of the Static Single Assignment (SSA) representation. This conversion not only simplifies stack instructions but also elucidates the behavior of the contract. Subsequently, the function signature is extracted from the memory stack through various jump operations. The selection of the function signature directly corresponds to the hash ID present on the stack. During the calldata formation phase, the EVM memory stack stores intermediate values and data from other execution sequences. Distinguishing between static and dynamic callee function parameters may impact the process, leading to variations in EVM generation protocols. The unique eight-character representation of the function signature is readily distinguishable within the EVM stack. Our approach to converting SSA-opcodes is designed to simplify the process of identifying a compatible operand that matches the designated function signature. This innovative process allows us to determine the hash ID within the context denoted as $\mathbb{B}$.

\begin{algorithm}[t]
\small
\SetAlgoNoLine
\normalem
    \caption{CFGs Connection For Gathering Data Paths}\label{algo:getRCFG}
  \SetKwData{Left}{left}\SetKwData{This}{this}\SetKwData{Up}{up}
  \SetKwProg{Fn}{Function}{:}{}\SetKwFunction{FindCALL}{FindCALL}\SetKwFunction{Find}{FindCALL}
  \SetKwFunction{getRCFG}{getRCFG} 
  \SetKwInOut{Input}{Input}\SetKwInOut{Output}{Output}
        % \DontPrintSemicolon
%   \setstretch{1.2}
	\Input{the caller bytecode contract $contract_t$;\\
           the callee bytecode contract $contract_c$
          }
	\Output{the connected CFG $rCFG$.}
	\BlankLine
    $CFG_t \leftarrow$ getCFG($contract_t$)\;
    $CFG_c \leftarrow$ getCFG($contract_c$)\; 
    $F_t \leftarrow$ getFuncPaths($CFG_t$); \Comment{get the set of function paths}\;
    $F_c \leftarrow$ getFuncPaths\&Sigs($CFG_c$); \Comment{get the set of function paths $f_c$ and signatures $sig_c$}\;
    rCFG $\leftarrow CFG_t$
    
    \For{$f_t \in F_t$}{ 
        \If{\text{isExistCALLs}$(f_t)$}{
            $f_p$, $f_n\leftarrow$ \text{SplitCALL}$(f_t)$;  \Comment{functions separation}\;
            $\mathbb{A}, \mathbb{B} \leftarrow f_p .\perp, f_n .\perp$; \Comment{path validation}\;
        }
    }
    \For{$f_{op} \in \mathbb{A}$}{
        \If{\text{isExistSig}$(f_{op}, sig_c)$}{
            \If{\text{isExistReturn}$(f_c)$}{
                ConnectReachableFunc($\mathbb{A}, \mathbb{B}, f_c$)\;
                RemoveUselessConnection($\mathbb{A},\mathbb{B}$)\;
                }
        }
    }
    \KwRet rCFG \;
% }
\end{algorithm}

Our investigation focuses on identifying specific contracts involved in transactions that occur at similar timestamps within DeFi protocol events. Within our analysis, in Algorithm \ref{algo:getRCFG}, we identify the attack contract as the caller in the attack sequence, while the targeted victim contract assumes the callee contracts. 

The algorithm makes a prior judgment on lines 5-10, which obtains all function paths that may exit external calls. We cut functions that involve calling operations into distinct code slices., and store them in $\mathbb{A}$ and $\mathbb{B}$, respectively. Then, the function block subgraph containing the precise function signature in the callee is connected to the $CFG_t$. In line 11, we derive the function signature mapped to the called function, thereby incorporating it as a distinct graph node within the $CFG_t$. The final phase of our methodology entails the amalgamation of the derived CFG. 

Subsequently, in lines 13-20, we generate distinct actions depending on the presence or absence of the \texttt{RETURN}. When \texttt{RETURN} exists, we execute the process of connecting external blocks in a manner that parallels the increment of nodes within the graph structure. Precisely, this entails the following steps. Initially, the deletion of the edge linking the function node $f_p$ and its counterpart $f_n$ is performed. Subsequently, a connection is established between the external function and $f_n$. Conversely, when the \texttt{RETURN} is absent, a slightly different procedure ensues. In this scenario, we delete the edge between $f_p$ and $f_n$, and the edge connecting the external function to $f_p$. This sequence of actions is concluded by the edges connection between the external function and $f_p$.

\noindent\textbf{Implementation.} The related CFGs construction contains three parts, \whiteding{1} getting CFG from a single contract, \whiteding{2} splitting contracts into basic blocks, and \whiteding{3} reconnecting the basic blocks. To achieve the first part \whiteding{1}, we construct the CFG from one contract by using the evm-cfg-builder module~\cite{evmcfgbuilder2024}, which has been explored by many works~\cite{Ethersplay2024,mossberg2019manticore,TrailofBits2024,ferreira2022elysium}. Then, as for the second part \whiteding{2}, we partition the functions and their corresponding function signatures according to Algorithm \ref{algo:getRCFG}. As a result, different basic blocks are divided by the CALL operation in the positioning function. Regarding the third part \whiteding{3}, we connect all reachable paths based on the collected function signatures, getting connected CFGs of two contracts.

\subsection{Data Path Selection}
\label{subsec::dps}
\noindent\textbf{Motivation. }When analyzing a CFG, the Depth-First Search (DFS) algorithm is utilized to select the data path that embeds as the contract feature. Note that the DFS path (i.e., the path selected with the DFS) could not be exactly equivalent to the real data path. Since functions can be called at different entries~\cite{ferreira2022elysium}, resulting in various data paths that are different from the DFS path. Meanwhile, in \cref{subsec::rcc}, we employ function calls to establish the CFG connection, which can cause the wrong connection.
Therefore, to address the problem, a path validation analysis will be conducted to obtain a more precise data path. After verification, we can identify a precise data path that aligns with the control flow by establishing edges.

As for the symbolic execution technique, we provide a detailed utilization within our framework. Given a data path $S_\pi$ = $\{s_0, s_1,..., s_n\}$, we define a function $P_v$ to determine the feasibility of the control flow path, which makes a judgment whether it can be executed normally from $s_0$ to $s_n$. The primary objective of function $P_v$ is to identify and select all data paths that can be accessed within the graph. Additionally, since we just symbolize the calculation representation to record the height of the symbolic stack, the path explosion issue is also resolved. Thus, there is a scarcity of space-consuming symbols when encountering loops. 

A data flow path $DP$ remains feasible, when the transfer condition of control flows on any adjacent blocks is positive. To determine the reachability of a given path, we introduce the $\Gamma(S_\pi)$ in the equation \ref{eqa:pathValidate}, which is a Boolean function to judge the feasibility of the path, the same capability with $\perp$ in the algorithm \ref{algo:getRCFG}. 

\begin{equation}
   \Gamma(S_\pi)=\bigwedge_{t=0}^{n-1} \bigvee_{DP \in CF(s_i, s_{i+1})} \bigwedge_{e \in CE(DP)} \Gamma_e(e)
\label{eqa:pathValidate}
\end{equation}

If the $\Gamma(S_\pi)$ returns \texttt{true}, the $S_\pi$ is a true branch, and the ! $\Gamma(S_\pi)$ is the false branch. We define $CF(s_i, s_{i+1})$ to represent the edges on the basic block $s_i$ to $s_{i+1}$ of control flow. Furthermore, $CE(DP)$ encompasses all the control edges within the data path $DP$, and $\Gamma_e(e)$ signifies the condition at the edge $e$. The edge is not directed by \texttt{JUMPI}, i.e., unconditional jumping, which defaults to true.

\begin{figure}[ht]
\centering
\includegraphics[width=0.85\linewidth]{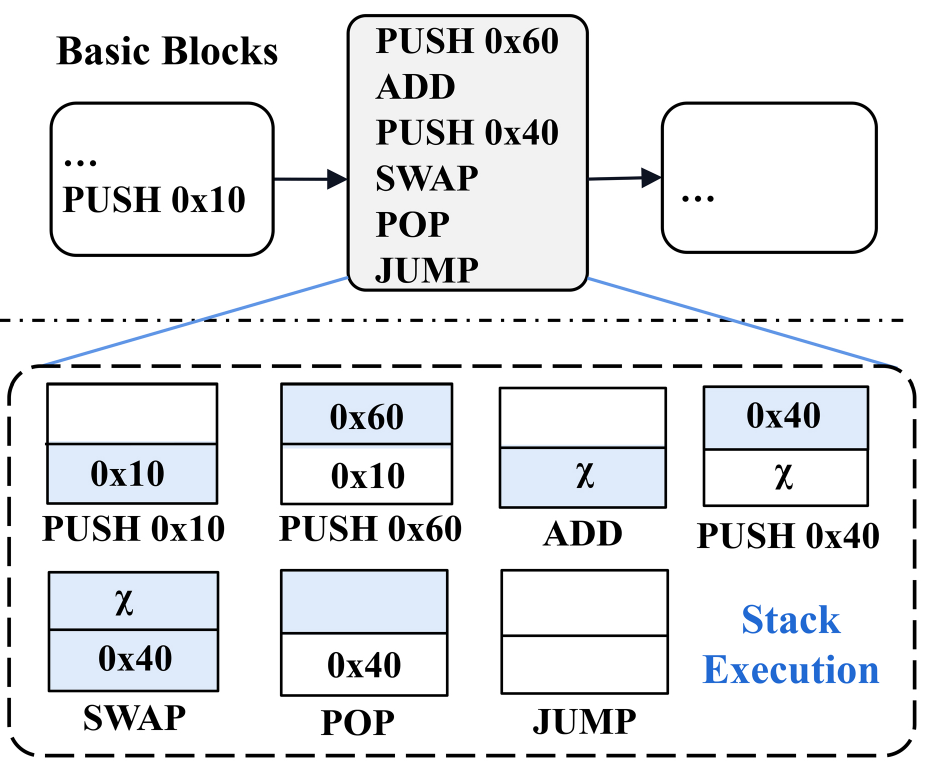}
\caption{An Example of Symbolic Stack Execution. $\chi$ represents the placeholder of the calculation result, and the blue rectangles mean the operands in or out of the stack for the current step.}
\label{fig:Symbolic_execution}
\end{figure}

Figure \ref{fig:Symbolic_execution} depicts a running example of the symbolic stack execution within a function. During this procedure, our primary focus is on the height of the stack, particularly the existence of a target signature during a jumping or calling operation. The evaluation of \texttt{ADD} operation results is circumvented by utilizing placeholders to maintain the stack's height. The function signature data includes various operations at the memory stack level, such as \texttt{PUSH}, \texttt{DUP}, \texttt{SWAP}, \texttt{AND}, and \texttt{POP}. We utilize symbolic execution methodology to manipulate the stack height associated with opcodes, thus allowing us to determine the stack's state during jumping.

In the data paths, we execute each operation in order, and we analyze the symbol stack's height in the last opcode of functions before jumping or calling. In the event that the target opcode lacks sufficient stack height at the moment of its occurrence, it is classified as an infeasible path. In our implementation, each split block is considered a node, and an edge shows the execution direction between blocks. The infeasible path will be removed if the execution of path operations results in an abnormal symbol stack height.

\begin{figure*}[!t]
\centering
\includegraphics[width=\linewidth]{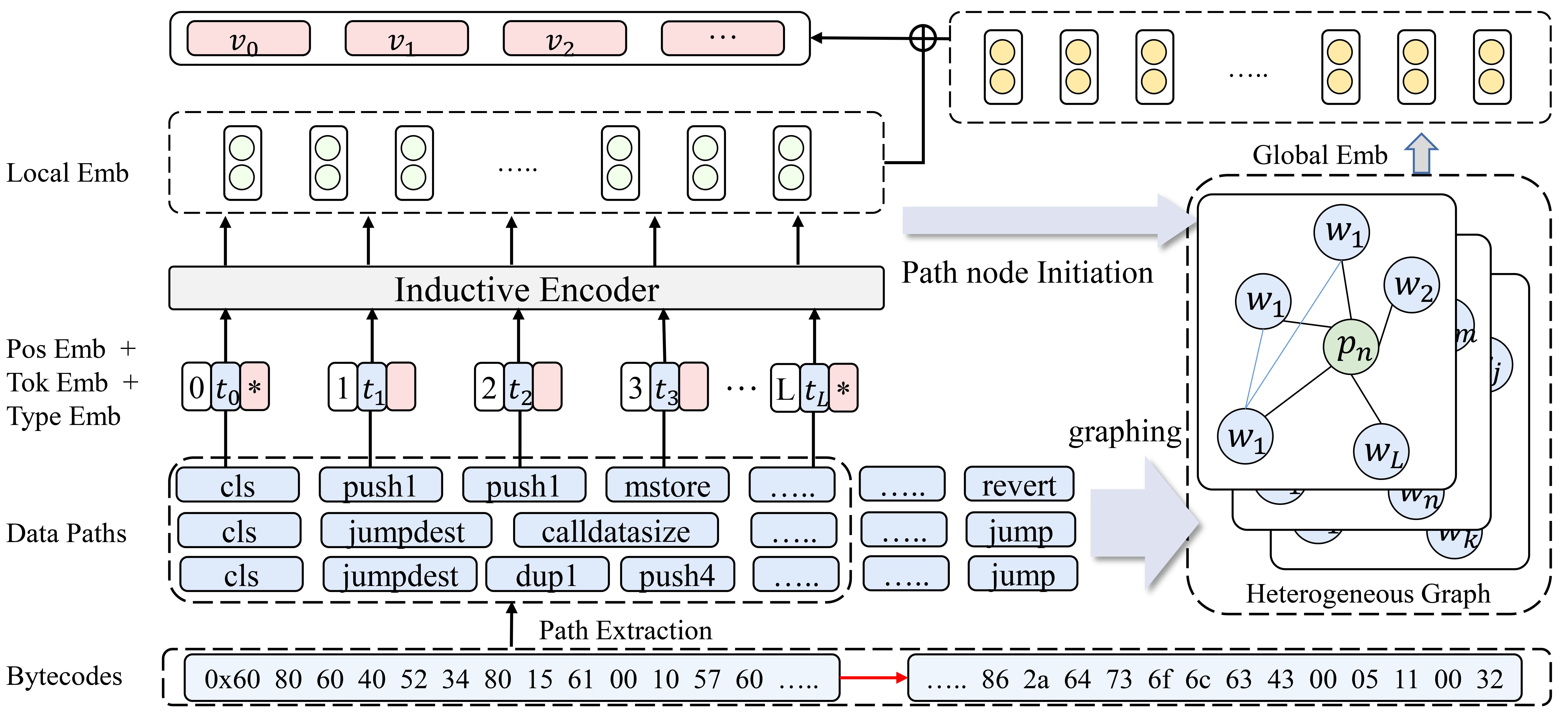}
\caption{The Encoder of Data Path Classification Model.}
\label{fig:model}
\vspace{-2ex}
\end{figure*}

\subsection{Model}
\label{subsec::model}
\noindent\textbf{Overview. }The model contains three parts as depicted in Figure \ref{fig:model}. Upon processing the two related contract bytecodes through \cref{subsec::rcc} and \cref{subsec::dps}, the resulting data path is utilized as input for our model. In the first part, we employ the inductive encoder to capture sequence features of the data paths. In the second part, we establish a heterogeneous graph structure for the data paths, treating each data path and opcode as individual nodes, and then learn the graph features. Finally, we merge the features obtained from the previous steps and utilize them for classification in the third part.

The training of the detection model utilized in DeFi protocols necessitates labeled data and the inclusion of all validated paths, encompassing both vulnerable DeFi protocols and secure instances. The model utilizes each validated data path and its corresponding sequence of instructions to learn about various DeFi attack patterns. These paths consist of EVM-compatible contracts in bytecode format and have been verified for feasibility (§ III-B). It first encodes the validated data paths pertaining to each bytecode contract into vectors Vs = $\{\pi_1, \pi_2, ..., \pi_n\}$, and subsequently inputs these vectors into the encoder module to extract hidden features. Finally, a linear and a softmax layer is deployed to aggregate all operations into a fixed-dimensional vector. This vector serves as the foundation for distinguishing whether the protocol is vulnerable or not.

\noindent\textbf{Proposed Architecture Description. }Figure \ref{fig:model} illustrates the encoder for processing the data flow paths. The \texttt{Pos Emb} means the position embedding, representing the position of tokens in the input sequence. The \texttt{Tok Emb} indicates the token embedding, representing the actual tokenized input data. The \texttt{Type Emb} represents the different types of tokens in each data path. Specifically, the * in Figure \ref{fig:model} marks the start point and end point of the data path, while the blank signifies the content between these points. Moreover, the \texttt{Pos Emb}, \texttt{Tok Emb}, and \texttt{Type Emb} are the same as BERT~\cite{kenton2019bert}. The heterogeneous graph in Figure \ref{fig:model} can be defined as $G=(N,T,E)$, where $N$ denotes different kinds of nodes, and each node $n_i \in N$ has corresponding type $T_i \in T$; $T$ is a type set of nodes, including opcode type $w$ and data path type $p$;  $E$ represents the set of edges connecting nodes.
A data path embedding $DP_\pi \in Vs$, which comprises a sequence of $n$ opcode embeddings represented as $\{t_1, t_2, ..., t_n\}$ in Figure \ref{fig:model}. Currently, relevant inductive learning methods, such as the transformer, truncate each input to a fixed dimension. Nevertheless, contracts deployed in DeFi entail intricate logic, resulting in generated data paths that are sizeable and use significant computational resources and memory. Consequently, we establish a heterogeneous graph within our corpus to acquire all features ultimately. While the graph structure captures overarching information, it needs to be improved in accessing important details of opcode sequence. Therefore, both inductive (e.g., BERT~\cite{kenton2019bert}, FastText~\cite{joulin2016fasttext}) and transductive (e.g., GNN~\cite{wu2020comprehensive}, GCN~\cite{yao2019graph}) methods are used to proficiently address DeFi contract scenarios, encompassing both position and global graph feature, are utilized.

In the process of constructing a graphical representation from a given data path, we employ the path sequence $DP_\pi$ to create a node connection matrix. The matrix, as depicted in Figure \ref{fig:model}, forms a heterogeneous graph. The black lines connect the path nodes and the corresponding opcode nodes, while the blue lines establish connections among the opcode nodes. 
We adopt established methodologies to establish these node connections as previously documented in works ~\cite{yao2019graph,lin2021bertgcn}. Specifically, we utilize the Pointwise Mutual Information (PPMI) and Term Frequency-Inverse Document Frequency (TF-IDF) techniques to determine the weights assigned to edges connecting opcodes and the path-opcode relationships.
PPMI, a statistical metric, quantifies the correlation between transactions and is calculated as $PPMI(i,j) = log({p(i,j)}/{p(i)p(j)})$, where $p(i,j)$ represents the probability of co-occurrence of elements $i$ and $j$. TF-IDF scores clarify the significance of a word within a document, with higher scores signifying a more significant influence.

\begin{equation}
   A_{ij} = f(i,j)=\begin{cases} PPMI(i,j) & \text{$i$, $j$ are opcodes} \\ TF-IDF(i,j)& \text{$i$ is path, $j$ is opcode}\\ 1 & \text{$i$ = $j$} \\ 0 & \text{otherwise}  \end{cases}
\label{eqa:Aij}
\end{equation}

Initially, we construct a matrix $X^a$ with dimensions of $(n_{path}+n_{opcode})\times d$, wherein $n_{path}$ refers to the path node, $n_{opcode}$ pertains to the opcode node, and $d$ denotes the dimension of the feature embedding. Based on Equation \ref{eqa:Aij}, the weight value is $PPMI(i,j)$ if the two nodes $i$ and $j$ are opcode nodes. In the scenario where $i$ is a path node and $j$ is an opcode node, the edge's weight value is deemed as $TF$-$IDF(i,j)$. If $i$ equals to $j$, then we assign it a weight of 1. Otherwise, it is considered as 0. After completing all traversals, $X^a$ is converted into a triangular matrix to hold the edge weights.

With the weight matrix generation method, we discovered that the section representing the path-path on matrix $X^a$ is 0. It means that the features of the path node itself cannot be learned. To solve it, we utilized the inductive learning method to learn the path feature vector from the opcode sequence and then initialized it into the graph. In this step, we truncate the input path to a fixed length and transform it into a one-hot encoding. Then, we convert it to a vector $X^b$ of dimensions ${(n_{path}+n_{opcode})\times d}$, aligning the heterogeneous graph vector. The corresponding dimension of opcode to 0, eliminating the influence of opcode features on graph nodes, i.e., $X^b=\dbinom{X^a_{path}}{0}$. Several studies~\cite{wu2021peculiar,wang2024contractcheck,zeng2023high,hu2023bert4eth,fang2023isyn} have demonstrated that BERT models can be fine-tuned to capture specific sequence patterns or behaviors for detecting vulnerabilities in smart contracts. So, we select the BERT structure, which sets 4 layers of transformer encoder, as the inductive encoder to obtain the feature vector $f_{X^b}$ of each path $DP_\pi = \{t_1, t_2, ..., t_n\}$, which is used to initiate the path nodes in $X^a$.  Simultaneously, we pass the path features through a linear and a softmax layer, demonstrated in Equation \ref{eqa:bert_classification}, to obtain the probability distribution. Note that we omit the bias to simply the explanation.

\begin{equation}
   Y_{bert} = softmax(Wf_{X^b})
\label{eqa:bert_classification}
\end{equation}

Once the path node initiation is completed on the heterogeneous graph $X^g$, we utilize a graph convolutional network (GCN) to extract features from all nodes. The GCN generates an output feature $X^{gcn}$ based on Equation \ref{eqa:GCN_matrix}. We then feed $X^{gcn}$ into a softmax layer to obtain the classification results.

\begin{equation}
   X^{gcn} = \sigma(\widetilde{D}^{-\frac{1}{2}}\widetilde{A}\widetilde{D}^{-\frac{1}{2}}X^gW)
\label{eqa:GCN_matrix}
\end{equation}
\begin{equation}
   Y_{gcn} = softmax(X^{gcn})
\label{eqa:output}
\end{equation}

The $\sigma$ is an activation function, the $\widetilde{D} = \sum_j\widetilde{A}_{ij}$ means the degree of the node $i$, and the $\widetilde{A}$ is an adjacency matrix that is constructed by the Equation \ref{eqa:Aij}. Thus, the $\widetilde{D}^{-\frac{1}{2}}\widetilde{A}\widetilde{D}^{-\frac{1}{2}}$ makes the $\widetilde{A}$ normalization. The $W \in \mathbb{R}^{C \times F}$ indicates a trainable parameters matrix, where $C$ expresses the dimension of the node feature vector, and $F$ represents the output dimension.

\begin{equation}
    Y = \lambda Y_{gcn} +(1-\lambda) Y_{bert}
\label{eqa:linearMerge}
\end{equation}

Subsequently, a linear interpolation is utilized to merge the inductive and transductive learning results, i.e., the $Y_{bert}$ and $Y_{gcn}$ in Equation \ref{eqa:linearMerge} are utilized to obtain the final result $Y$. The $\lambda \in [0, 1]$ controls the balance of two probability distributions. 

\begin{equation}
    loss(x,i) = W_i(-x_i +log(\sum_{j} exp(x_j))),
\label{eqa:crossentropy}
\end{equation}
% \vspace{-2ex}
% Moreover, 
where we simply employ the cross entropy loss function $loss(x, i)$ at Equation \ref{eqa:crossentropy} with weight distributions $W=[W_1, W_2]$ throughout the training, validation, and testing. For the $i_{th}$ category, the $W_i$ is calculated as the total number of categories divided by the number of the $i_{th}$ class, and $x=(x_0,x_1)$ is a non-softmax output. Finally, calculating the weighted sum of the $loss(x, i)$ generated by the two classes, and dividing the sum by the number of classes to obtain the final weighted-cross entropy loss.

%% file: sections/evaluation.tex
\section{Results}
\label{sec::evaluation}

In this section, we evaluate the performance of DeFiTail in detecting compromised projects within real-world DeFi protocols at the bytecode level. Specifically, we address the following questions:
\begin{itemize}
    \item[RQ1] \textbf{Can DeFiTail outperform other vulnerability detection tools?} 
    We aim to investigate DeFiTail's capability in detecting real DeFi protocols, and how the efficiency compares to state-of-the-art (SOTA) tools.
    
    \item[RQ2] \textbf{How do distinct settings impact the efficiency of DeFiTail?}
    We aim to examine the individual impact of DeFiTail's each component, and how the effectiveness of related CFG construction and path selection.

    \item[RQ3] \textbf{Can our method identify attacks in the real world?}
    We aim to explore DeFiTail's capability to be applied, or to monitor real-world attacks effectively.
    
    \item[RQ4] \textbf{Can DeFiTail inspect real-world vulnerabilities?}
    We aim to assess the adaptability of DeFiTail, investigating its capacity to adjust to real-world environments.

\end{itemize}

\subsection{Experiment Settings}
\label{subsec::experiment_setting}

\subsubsection{Dataset}
To evaluate the efficacy of the DeFiTail, we meticulously curated a dataset to facilitate comprehensive comparative evaluations. This dataset comprises a 3,216 substantial collection of 14,301 data paths extracted from 3,216 instances of hacked DeFi events. The data was meticulously gathered from the REKT database~\cite{REKT-database2023solidity}, one of the largest crypto databases containing DeFi scams, hacks, and exploits. The dataset collection process consists of a total of three filtering criteria. First, we only collect data from blockchains that are compatible with EVM. Second, we only collect events where the loss amount exceeds \$10,000. Third, we only collect data with detailed audit reports or attack transaction execution traces, and it must contain the relevant contract address or code. Each incident analysis within our dataset entails an exhaustive examination of malicious accounts and compromised token contracts based on abnormal transactions. Moreover, various attack types are encompassed, including access control, flash loan exploits, honeypots, and rug pull incidents. Significantly, these malicious activities transpired across 25 EVM-compatible blockchains, such as Ethereum, Binance, EOS, Polygon, Arbitrum, Fantom, et al. The dataset is then divided into training and test sets in a 90\%: 10\% ratio for evaluation. The positive label numbers for access control and flash loan exploitation are 61 and 89. As detailed in Table \ref{tab:dataset_statistics}, access control and flash loan exploits jointly constitute a mere 4.66\% and 5.60\% of the entire dataset. Recognizing this imbalance, we implemented an oversampling strategy to rebalance the training set, ensuring adequate exposure to the minority classes.

\begin{table}[ht]
\small
\centering
  \caption{The Dataset Statistics.}
  \label{tab:dataset_statistics}
  \begin{tabular}{c  c  c }
    \toprule
    Category & Ratio (\%) & Avg Size (byte) \\
    \midrule
    Access Control &  4.66 & 17,349.16 \\
    Flash Loan Exploits &  5.60  & 25,641.41 \\
  \bottomrule
  % \vspace{-4ex}
\end{tabular}
\end{table}

\subsubsection{Metric and Environment}
Since we do not have the ground truth dataset, the presence of a contract solely within a class cannot be definitively determined. Therefore, under our consideration, accuracy is employed as the criterion of evaluation. For instance, consider the scenarios where contract $\alpha$ within a DeFi protocol is revealed to include access control patterns. However, the flash loan exploit that exists within $\alpha$ remains uncertain. Therefore, for a negative sample, we do not know whether access control or flash loan exists on it. Thus, our primary emphasis lies in determining the efficacy of accurately identifying potential threats. All experiments are conducted on a server with an NVIDIA GeForce GTX 4070 Ti GPU and an Intel(R) Core(TM) i7-12700 CPU. The parameter amount of the inductive encoder (i.e., BERT) we utilized is around 15 million. Regarding graph learning, we use a 2-layer GCN with 200 hidden channels. In addition, we added a dropout of 0.5 to the model to prevent overfitting, which discards 50\% of the neurons during the training process.

\subsection{Comparison Analysis (RQ1)}
\label{subsec::comparison}

Table \ref{tab:comparison_results_interaction} illustrates how DeFiTail can be used to identify and prevent security vulnerabilities in smart contracts, using DeFiTail to detect access control and flash loan exploits. In line with DeFiTail's design objectives, these vulnerabilities arise from the interaction of different contracts. In terms of access control, a contract may be vulnerable to malicious behavior if there has been unintentional manipulation of the contract's permissions. Unauthorized permissions may potentially be granted to other contracts, leading to instigation. Additionally, users have the option to employ the flash loan mechanism to heighten the liquidity of funds, ultimately augmenting conversion between different assets. Given that this process involves many token contracts, Flash Loan incidents may occur when interacting with contracts.

\begin{table}[ht]
\small
\centering
  \caption{The Comparison Results in Different Tools}
  \label{tab:comparison_results_interaction}
  \begin{tabular}{c c c }
    \toprule
    \textbf{Model Name} & \textbf{Access Control }& \textbf{Flash Loan Exploits} \\
    \midrule
    Slither\cite{feist2019slither}  & 18.67\%  & -- \\
    Maian\cite{nikolic2018finding}    & 45.65\%  & -- \\
    Mythril\cite{mythril2019} & 64.91\% & --\\
    Ethainter\cite{brent2020ethainter} & 42.86\% &--\\
    % SPCON\cite{liu2022finding} & &-- \\
    Achecker\cite{ghaleb2023achecker} &81.82\% & --\\
    Midnight\cite{ramezany2023midnight} &-- & 86.17\% \\
    DeFiTail & 98.39\%& 97.43\%\\ 
  \bottomrule
  % \vspace{-4ex}
\end{tabular}
\end{table}

To demonstrate the effectiveness of our framework, we undertake a comparative analysis with SOTAs \cite{mythril2019,brent2020ethainter,ghaleb2023achecker,ramezany2023midnight}. As for access control, we conducted a comparison analysis of access control detectors on a single device, including Slither~\cite{feist2019slither}, Maian~\cite{nikolic2018finding}, Mythril~\cite{mythril2019}, Ethainter~\cite{brent2020ethainter}, and Achecker~\cite{ghaleb2023achecker}. Maian~\cite{nikolic2018finding} assesses prodigal, suicidal, and greedy contracts, which predominantly concentrates on access control issues arising from token transfers. Slither~\cite{feist2019slither} utilizes a static analysis approach to identify specific operational behaviors, revealing novel behaviors that can bypass permissions. Mythril~\cite{mythril2019} uses symbolic execution techniques to analyze whether a given input in a program may reach a vulnerable state. Specifically, it uses an SMT (Z3) solver to find out whether a program satisfies the constraints of a vulnerable state. Ethainter~\cite{brent2020ethainter} uses Datalog to store information flows, and then uses data filtering technology to check modifier-related transactions. Achecker~\cite{ghaleb2023achecker} analyzes data dependence through input and state variables, checking conditions of permission control. When using Mythril, specific related vulnerabilities are deemed relevant as access control is not directly detected. In the situations of real DeFi scams, the accuracy in the table will be lower than 64.91\%. Since the Ethainter analyzes contracts on three different blockchains, we detected all avaliable contracts and calculated an accuracy of 42.86\%. Note that the Ethainter provides a public website\footnote{https://library.dedaub.com/}, the comparison results in Table \ref{tab:comparison_results_interaction} are from it. The Achecker, which is a tool that is commonly utilized to identify permission vulnerabilities. After testing, Achecker achieved an accuracy of 81.82\%. DeFiTail achieved 98.39\% accuracy in the test set, outperforming other SOTA tools in Table \ref{tab:comparison_results_interaction}.

As for flash loan exploits, there is limited research on it~\cite{wang2021towards,qin2021attacking,EEAtools2023quantstamp,ramezany2023midnight}, and the only open-source tool we found is Midnight~\cite{ramezany2023midnight}, which is a real-time tool based on event analysis. Midnight associates events with contracts based on real-time transactions, so we obtained the transactions of the test set. Finally, we compare the results of DeFiTail and Midnight in Table \ref{tab:comparison_results_interaction}.

\begin{tcolorbox}[boxrule=1pt,boxsep=1pt,left=3pt,right=3pt,top=3pt,bottom=3pt]
\textbf{Answer to RQ1.}
DeFiTail exhibits remarkable efficiency in the detection of access control and flash loan exploits, achieving accuracies of 98.39\% and 97.43\%, respectively.
\end{tcolorbox}

\subsection{Ablation Analysis (RQ2)}
\label{subsec::ablation}

Within DeFiTail, we have evaluated the individual effects of data path selection and CFG connection, and the results are presented in Table \ref{tab:ablation_analysis}.
% , where we evaluated the individual effects of data path selection and CFG connection. 
Furthermore, the influence of Transformer and heterogeneous graph was investigated. 

DeFiTail performs 0.48\% and 41.38\% for access control better with path selection and related CFGs connection than without them. The detection accuracy of access control attack events is surprisingly improved by path selection. However, the detection performance experiences a significant decline in the absence of a relevant contract to detect the vulnerability jointly. The occurrence arises due to the vulnerability relying heavily on the interaction of different contracts, and it is a critical operation to invoke the malicious contract from other contracts to obtain permissions.

Contrary to access control, models without path related CFG connections perform better than those without. Furthermore, for flash loan events, the results are reversed. The detection of flash loan events is negatively affected by only connecting two related contracts without a path selection process. On the other hand, DeFiTail-np, which only includes the construction of related CFGs, is significantly lower at 12.24\% compared to DeFiTail. Due to the minor difference of only 0.21\% between DeFiTail and DeFiTail-npc, which lacks both features, it suggests that the connection of two related contracts will enhance the detection results. Therefore, it is evident that the path feasibility verification when connecting related CFGs has a positive impact on the detection results.

\begin{table}[ht]
\small
\centering
  \caption{The Ablation Analysis Results. np/npc means the model without path selection/rCFG construction. BERT represents the model with the pre-trained BERT-base.}

  \label{tab:ablation_analysis}
  \begin{tabular}{c c c }
    \toprule
    \textbf{Model Name} & \textbf{Access Control} & \textbf{Flash Loan Exploits}\\
    \midrule
    DeFiTail-np         & 98.23\%   & 85.19\%  \\
    DeFiTail-npc        & 57.33\%   & 97.22\%  \\
    \hline
    DeFiTail-LSTM       & 88.89\%   & 95.00\%  \\
    DeFiTail-BERT       & 89.71\%   & 96.30\%  \\
    \textbf{DeFiTail}   & 98.39\%   & 97.43\%  \\
  \bottomrule
\end{tabular}
\end{table}
% \vspace{-2ex}

In our framework, we incorporate path selection, related CFG construction, and contract embedding to extract localized features during our model design process. Additionally, we construct a heterogeneous graph to capture global features across all contracts. However, we have found that directly utilizing pre-trained models, such as BERT, for classification tasks has been effective. Consequently, we have conducted experiments displayed in Table \ref{tab:ablation_analysis}. The LSTM and pre-trained BERT-base models are integrated into our framework, to derive the final classification results in these experiments. 
Specifically, due to the design of our graph, the two kinds of nodes can be represented. So the graph feature we extracted is the heterogeneous graph feature that captures the distinct features of both opcodes and data paths. Thus, the features acquired by GCN inherently consist of heterogeneous graph features. We then compare these results with the complete DeFiTail approach, which includes one additional global feature beyond DeFiTail-BERT. The results show that access control detection was influenced by different components (LSTM and BERT) and global graph features, which significantly impacted access control detection. However, there is a slight discrepancy in the effectiveness of the three detection techniques in identifying flash loan exploits.

\begin{table}[ht]
\small
\centering
  \caption{The Results in Signal-Contract Scams Detection.}
  \label{tab:ablation_analysis1}
  \begin{tabular}{c c}
    \toprule
    \textbf{Model Name}     & \textbf{Accuracy(\%)}\\
    \midrule
    DeFtTail-HoneyPot       & 60.39 \\
    DeFiTail-RugPull        & 72.51 \\
  \bottomrule
\end{tabular}
\end{table}
% \vspace{0.2em}

To further emphasize the limited influence of detecting DeFi events without interaction information, we conducted experiments as outlined in Table \ref{tab:ablation_analysis1}. In the experimental result, we performed CFG construction on a single token contract in honeypot and rug pull scams. Subsequently, we validated the path inaccessibility, followed by model training and experimental testing. The detection accuracy of 60.39\% and 72.51\% is significantly lower than that of CFG connections, i.e., access control and flash loan exploits.

\begin{tcolorbox}[boxrule=1pt,boxsep=1pt,left=3pt,right=3pt,top=3pt,bottom=3pt]
\textbf{Answer to RQ2: } The presence of both path selection and related CFG connection in access control and flash loan exploits is crucial for the dependencies of DeFiTail.
\end{tcolorbox}

\subsection{Detection Capability (RQ3)}
\label{subsec::case}
\begin{table}[h]
\small
\centering
% \vspace{-3ex}
\renewcommand{\arraystretch}{1}
  \caption{The Evaluation Results on CVE Dataset. N/A means that necessary information is missing.}
  \label{tab:case_analysis}
\setlength{\tabcolsep}{3mm}{
  \begin{tabular}{c c c c c c c c}
    \toprule
    \textbf{CVE-ID} & \rotatebox{90}{\textbf{Slither}} &  \rotatebox{90}{\textbf{Maian}} & \rotatebox{90}{\textbf{Mythril}} & \rotatebox{90}{\textbf{SPCon}} & \rotatebox{90}{\textbf{AChecker}}  & \rotatebox{90}{\textbf{DeFiTail}}\\
    \midrule
    CVE-2018-10666 & \scalebox{0.75}{\usym{2613}}  & \scalebox{0.75}{\usym{2613}} & \scalebox{0.75}{\usym{2613}}  & $\checkmark$  & $\checkmark$  & $\checkmark$  \\
    CVE-2018-10705 & \scalebox{0.75}{\usym{2613}}  & \scalebox{0.75}{\usym{2613}} & \scalebox{0.75}{\usym{2613}}  & $\checkmark$  & $\checkmark$  & N/A  \\
    CVE-2018-11329 & \scalebox{0.75}{\usym{2613}}  & \scalebox{0.75}{\usym{2613}} & \scalebox{0.75}{\usym{2613}}  & $\checkmark$  & $\checkmark$ & \scalebox{0.75}{\usym{2613}} \\
    CVE-2018-19830 & \scalebox{0.75}{\usym{2613}}  & \scalebox{0.75}{\usym{2613}} & $\checkmark$   & N/A & $\checkmark$ & $\checkmark$  \\
    CVE-2018-19831 & \scalebox{0.75}{\usym{2613}}  & \scalebox{0.75}{\usym{2613}} & \scalebox{0.75}{\usym{2613}}  & $\checkmark$  & $\checkmark$  & $\checkmark$  \\
    CVE-2018-19832 & \scalebox{0.75}{\usym{2613}} & $\checkmark$  & \scalebox{0.75}{\usym{2613}}  & $\checkmark$ & $\checkmark$  & $\checkmark$  \\
    CVE-2018-19833 & \scalebox{0.75}{\usym{2613}}  & \scalebox{0.75}{\usym{2613}} & \scalebox{0.75}{\usym{2613}}  & N/A & $\checkmark$ & $\checkmark$  \\
    CVE-2018-19834 & \scalebox{0.75}{\usym{2613}}  & \scalebox{0.75}{\usym{2613}} & \scalebox{0.75}{\usym{2613}}  & N/A & $\checkmark$ & $\checkmark$  \\
    CVE-2019-15078 & \scalebox{0.75}{\usym{2613}} & $\checkmark$ & $\checkmark$  & $\checkmark$ & $\checkmark$  & $\checkmark$  \\
    CVE-2019-15079 & \scalebox{0.75}{\usym{2613}}  & \scalebox{0.75}{\usym{2613}} & \scalebox{0.75}{\usym{2613}}  & $\checkmark$ & \scalebox{0.75}{\usym{2613}} & $\checkmark$  \\
    CVE-2019-15080 & \scalebox{0.75}{\usym{2613}}  & \scalebox{0.75}{\usym{2613}} & \scalebox{0.75}{\usym{2613}}  & $\checkmark$ & $\checkmark$  & $\checkmark$ \\
    CVE-2020-17753 & $\checkmark$  & \scalebox{0.75}{\usym{2613}} & $\checkmark$  & \scalebox{0.75}{\usym{2613}} & \scalebox{0.75}{\usym{2613}} & $\checkmark$  \\
    CVE-2020-35962 & \scalebox{0.75}{\usym{2613}}  & \scalebox{0.75}{\usym{2613}} & $\checkmark$   & \scalebox{0.75}{\usym{2613}} & \scalebox{0.75}{\usym{2613}} & $\checkmark$  \\
    CVE-2021-34272 & \scalebox{0.75}{\usym{2613}}  & \scalebox{0.75}{\usym{2613}} & \scalebox{0.75}{\usym{2613}}  & $\checkmark$ & $\checkmark$ & $\checkmark$  \\
    CVE-2021-34273 & \scalebox{0.75}{\usym{2613}}  & \scalebox{0.75}{\usym{2613}} & \scalebox{0.75}{\usym{2613}}  & \scalebox{0.75}{\usym{2613}} & $\checkmark$  & $\checkmark$  \\
  \bottomrule
\end{tabular}}
\vspace{-4ex}
\end{table}

\begin{table*}[!t]
\centering
\caption{Descriptions for the 5 Kinds of DeFi Exploits}
% \vspace{2pt}
\resizebox{\linewidth}{!}{
\begin{tabular}{|l|l|}
\hline
DeFi exploits &
  Description \\ \hline
\rowcolor{black!15} Repetition Abuse &
  \begin{tabular}[c]{@{}l@{}}
  The contract leverages the atomicity of flash loans to allow multiple function calls for\\ a short time, resulting in arbitrage from the liquidity pool.
  \end{tabular} \\ %
 Unsafe Unintended Exploit &
  \begin{tabular}[c]{@{}l@{}} External calls to an unintended address, leading to an unknown account disrupting the \\normal execution logic and gaining unauthorized access.\end{tabular} \\ %
\rowcolor{black!15} Signature Violated Exploit &
  \begin{tabular}[c]{@{}l@{}} Message calls invoke functions using a computed 4-byte function signature. When the \\targeted function does not exist, the failed call leads to unintended behavior.\end{tabular} \\ %
Insecure interfaces Exploit &
  \begin{tabular}[c]{@{}l@{}}When implementing ERC interfaces for DeFi protocols, the block information is utilized \\in an imprecise manner, resulting in the unauthorized appropriation of token privileges.\end{tabular} \\ %
  \hline
\rowcolor{black!15} Unrestricted Token Transfer &
  \begin{tabular}[c]{@{}l@{}}The transfer operation lacks restrictions on the token sender, allowing any individual to\\ withdraw tokens from the contract.\end{tabular} \\ %
  \hline
\end{tabular}
}
\label{table::defiexploits}
\vspace{-2ex}
\end{table*}

To verify the effectiveness of DeFiTail in preventing access control scams in the real environment, we evaluate Slither~\cite{feist2019slither}, Maian~\cite{nikolic2018finding}, Mythril, SPCon~\cite{liu2022finding}, AChecker, and DeFiTail on a CVE dataset listed in Table \ref{tab:case_analysis}, representing their ability to detect attacks in the real-world environment. Slither~\cite{feist2019slither} tracks variable and state changes in the AST structure to see if inappropriate vulnerability patterns occur. By simulating the execution process of the contract, Maian~\cite{nikolic2018finding} uses symbolic execution technology to establish and explore the state machine model to check the vulnerability of different states of the contract. 
SPCon~\cite{liu2022finding} identifies privileged functions that should not have been accessed by mining vulnerabilities in the transaction history.

Table \ref{tab:case_analysis} indicates that the tag "N/A" denotes inadequate input data necessary for analysis. For example, CVE-2018-19830, CVE-2018-19833, and CVE-2018-19834 are all designated as "N/A" on SPCon because of insufficient transaction data for historical role mining simulations related to access control or permission issues. Additionally, CVE-2018-10705 is marked "N/A" in the DeFiTail score because there are no associated contracts in the malicious events.

In addition, to ensure maximum detection, we performed vulnerability events using symbolic execution, i.e. Maian and Mythril, for at least 30 minutes on the cases analyzed in Table. \ref{tab:case_analysis}. Unfortunately, due to a lack of maintenance, we were originally unable to run the SPCon's source code to collect test results. Nevertheless, we have demonstrated the effectiveness of DeFiTail by comparing it to the detection method mentioned in the original paper. This method is considered to be the state-of-the-art tool for detecting real vulnerabilities.

\begin{tcolorbox}[boxrule=1pt,boxsep=1pt,left=3pt,right=3pt,top=3pt,bottom=3pt]
\textbf{Answer to RQ3: } DeFiTail exhibits a remarkable ability to identify security vulnerabilities compared to the state-of-the-art, successfully detecting 86.67\% of 15 CVE incidents.
\end{tcolorbox}

\subsection{Adaptability (RQ4)}
\label{subsec::adaptability}
To assess the adaptability of DeFiTail in real-world scenarios, we utilized it to monitor the operational contracts of DeFi protocols on the Ethereum blockchain. 

However, in the DeFi security environment, false positives can lead to legitimate protocols being incorrectly categorized as malicious activities. For instance, directly profiling by monitoring all accounts that engage with the protocol tends to elevate the false positive rate. Therefore, during the evaluation, we used a strategy aimed at optimizing the false positive rate. Given that transactions on Ethereum follow a power-law distribution—where a small number of accounts account for the majority of transactions—we opted to exclude the top accounts, as identified by Etherscan~\cite{Etherscan}, to mitigate false positives. While this approach may increase the false negative rate, it effectively conserves computing resources.

After a thorough inspection period of 60 hours, we detected several malicious exploits, and the contracts will be recorded in our online artifact~\cite{defitail2024artifact}. Since, to the best of our knowledge, none of the previous studies subdivide access control and flash loan exploitation, we refine the found malicious vulnerabilities into 5 categories as shown in Table. \ref{table::defiexploits}. Simultaneously, we report the exploits to the CVE repository and the manufacturer.

The initial categories identified in this study consist of two primary dimensions, i.e., access control and flash loan exploits. We conducted a comprehensive examination of the 500 verified contracts over a duration of 60 hours, during which we undertook a detailed manual classification of these contracts into the 5 distinct categories detailed in Table. \ref{table::defiexploits}. Specifically, we have reported a number of vulnerabilities, including repetition abuse (CVE-2024-51169), unsafe unintended exploit (CVE-2024-51167, CVE-2024-51172), signature violated exploit(CVE-2024-51170), insecure interfaces exploit (CVE-2024-51171, CVE-2024-51173, and CVE-2024-51174), and unrestricted token transfer (CVE-2024-51168).

Take the repetition abuse in the flash loan exploits as an example, flash loan exploit attacks leverage vulnerabilities in DeFi by borrowing assets from a flash loan agreement. As illustrated by the Oracle contract in Listing 1 and the attacker contract in Listing 2, the exploit extends through a sequence of actions. The attacker initiates the preparation operation by leveraging a flash loan mechanism, as shown in line 9 of Listing 2, to acquire a substantial quantity of stablecoins. Subsequently, during the attacking execution phase in line 14 of Listing 2, the attacker deposits the borrowed assets into Oracle’s liquidity pool. Then, the attacker follows an abrupt removal of a significant portion of the liquidity, purposely destabilizing the pool’s stability. The resulting imbalance induces volatility in the pool’s asset valuation metrics, thereby distorting the price of tokens in the pool. A critical exposure is exploited in line 21 of Listing 2, where the attacker uses the price lookup function \texttt{calculateRewards(address user)} in the victim as a view property, which makes the function not need to consume gas. Therefore, the attacker keeps querying the asset prices in the liquidity pool, and when the price in the liquidity pool is inflated and not updated, all the funds in the pool are extracted. Finally, the flash loan repayment is executed.

\noindent\textbf{Distinctions.} 
Previous flash loan attacks, exemplified by the Harvest Finance incident on October 26, 2020~\cite{chen2024flashsyn}, typically involved executing multiple cycles of flash loans to yield several profits through consecutive iterations(i.e., \textbf{repeated} cycles of flash loans with \textbf{one} profit). 
However, the situation described in Listing 2 highlights a strategy focused on acquiring small, repeated profits within a singular flash loan transaction(i.e., \textbf{one} cycle of a flash loan with \textbf{repeated} profits). This strategy seeks to optimize the automated realization of the maximum potential profit from a single flash loan cycle.

\begin{lstlisting}[language=Solidity, 
                    basicstyle=\fontsize{7}{8}\ttfamily,
                    numbers=left,
                    captionpos=b,
                    commentstyle=\color{orange},
                    aboveskip = 1em,
                    belowskip = 1em,
                    numbersep= -1em,
                    caption= The Simplified Snippets of an Oracle,
                    label=lst:oracle,
                    ] 
  contract Oracle {
    IPool public pool;  
    mapping(address => uint) public rewards;
    function calculateRewards(address user) external view returns(uint) {
      uint price = pool.get_virtual_price();  
        return rewards[user] * price; 
      }
    function claimRewards() external {
      uint amount = calculateRewards(msg.sender);
      payable(msg.sender).transfer(amount);  
    }
  }
    \end{lstlisting} 
    \vspace{-2ex}
    \begin{lstlisting}[language=Solidity, 
                    basicstyle=\fontsize{7}{8}\ttfamily,
                    numbers=left,
                    captionpos=b,
                    commentstyle=\color{orange},
                    aboveskip = 1em,
                    belowskip = 1em,
                    numbersep= -1em,
                    caption= The Simplified Snippets of the Attacker,
                    label=lst:B2X,
                    ] 
  contract FlashLoanAttacker {
    address pool;
    address victim;
    constructor(address _pool, address _victim) {
      pool = _pool;
      victim = _victim;
    }
    // Initiation
    function attack(uint256 loanAmount) external {
      require(msg.sender == owner, "Unauthorized");
      pool.flashLoan(loanAmount); 
    }
    // Attacking
    function executeOperation(uint256 amount) external {
      // 1. Liquidity Manipulation   
      IERC20(USDC).approve(pool, type(uint).max);
      IPool(pool).add_liquidity([amount, 0], 0);
      // 2. Price Fluctuation
      IPool(pool).remove_liquidity(amount*95/100, [0, 0]);
      // 3. Repetition Abuse
      for (uint i = 0; i < 5; ){
        currentReward = victim.calculateRewards(pool);
        if (currentReward > amount){ break; }
        i++;
      }
      // 4. Arbitrage
      victim.claimRewards();
      // 5. Repayment
      IERC20(pool).transfer(pool, amount*1001/1000); 
    }
  }
    \end{lstlisting} 

\vspace{-2ex}

\begin{tcolorbox}[boxrule=1pt,boxsep=1pt,left=3pt,right=3pt,top=3pt,bottom=3pt]
\textbf{Answer to RQ4: } With DeFiTail, 5 categories of exploits are found, proving its inspection capability in the real world.
\end{tcolorbox}

\section{Discussion}
\label{sec:discussion}

In this section, we will discuss the detected incident, threat to validity, and limitations. For the reason that we detect contracts in bytecodes that are difficult to understand, we demonstrate these in source code, which can improve clarity.

\subsection{Detected Incident}
DeFiTail can detect exploitation during the dynamic execution process, even though it uses static techniques.
Within the Listing \ref{lst:B2X}, there is an access control issue on line 4. The function \texttt{owned()} is declared public, allowing external accounts (i.e., contract accounts and EOAs) to execute it and gain access permissions. It gives the authorization of \texttt{owner}, which represents that external accounts can enter contracts or functions by utilizing the \texttt{modifier} on line 7, ultimately resulting in a breakdown in access control.

\begin{lstlisting}[language=Solidity, 
                    basicstyle=\fontsize{8}{9}\ttfamily,
                    numbers=left,
                    captionpos=b,
                    commentstyle=\color{orange},
                    aboveskip = 1em,
                    belowskip = 1em,
                    numbersep= -1em,
                    caption= The Simplified Snippets of BTC2X,
                    label=lst:B2X,
                    ] 
  contract Owned {
    address public owner;
    // The key point
    function owned() public {
        owner = msg.sender;
    }
    modifier onlyOwner {
        require(msg.sender == owner);
        _;
    }
    function transferOwnership(address newOwner) onlyOwner public {
        owner = newOwner;
    }
  }
\end{lstlisting} 

The attacker creates transactions that call functions that lack access control in order to change the contract creator state. As an example, the attack interaction process with the attacker contract $\delta[A_{att}]$ shown in Listing \ref{lst:B2X_attack}, contract $\delta[A_{own}]$ shown in Listing \ref{lst:B2X} is deployed at $A_{own}$, and the attacker's contract $\delta[A_{att}]$ is deployed at $A_{att}$. At first, the $\delta[A_{att}]$ is passed in the $A_{own}$. Then, transaction $T=\{A_{att}, A_{own}, owned()\}$ is created, before invoking the $attack()$ function in line 8 of $\delta[A_{att}]$. Then, the $A_{own}$ as the input parameter affects the $msg.sender$ in line 5 of $\delta[A_{own}]$, resulting in the $owner$ changes to attacker $A_{att}$ and controlling the permission. DeFiTail mainly focuses on the transaction process with sensitive data and their interaction styles.

\begin{lstlisting}[language=Solidity, 
                    basicstyle=\fontsize{8}{9}\ttfamily,
                    numbers=left,
                    captionpos=b,
                    commentstyle=\color{orange},
                    aboveskip = 1em,
                    belowskip = 1em,
                    numbersep= -1em,
                    caption= The Attack Snippets to BTC2X,
                    label=lst:B2X_attack,
                    ] 
  // Attacker
  contract Attack {
    Owned ownedContract;
    constructor(address _ownedContractAddr) {
        ownedContract = Owned(_ownedContractAddr);
    }
    // attack operation
    function attack() public {
        ownedContract.owned();
    }
    // transfer the ownership
    function takeOwnership(address newOwner) public {
        ownedContract.transferOwnership(newOwner);
    }
}
\end{lstlisting} 
\vspace{-2ex}

\begin{figure}[ht]
\centering
\includegraphics[width=.60\linewidth]{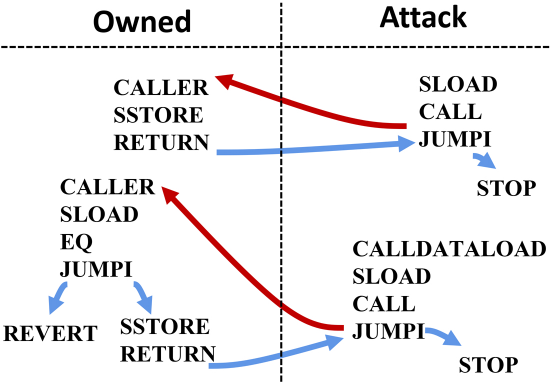}
\caption{The Simplified Schematic of the Opcode Flow.}
\label{fig:opcode_flow}
% \vspace{-2ex}
\end{figure}

For an attacker who wants to gain access to a victim's contract, it mainly consists of three steps. 

In the first step, the attacker executes \texttt{CALLDATALOAD} to read the constructor and executes \texttt{SSTORE} to store the victim's address into slot (i.e., line 5 of Listing \ref{lst:B2X_attack}).

Figure \ref{fig:opcode_flow} shows the interaction in the second and third steps.
In the second step, the attacker performs \texttt{STATICCALL} operation on line 8 of Listing \ref{lst:B2X_attack} to invoke the \texttt{owned()} function of line 4 in the victim contract Listing \ref{lst:B2X}. Then \texttt{SLOAD}(to get the slot of the owner variable), \texttt{CALLER}(to get the attack contract address), and \texttt{SSTORE}(to write the attack contract address to the slot of the owner variable) are executed.

In the third step, the \texttt{transferOwnership(address)} function in the victim's contract on line 12 of Listing \ref{lst:B2X_attack} is called by the attacker, and then the require verification of \texttt{JUMPI} jump to line 7 in Listing \ref{lst:B2X} is performed. This validates \texttt{SLOAD}(to read the value of the slot in the owner variable) and \texttt{EQ}(to verify that \texttt{CALLER} and owner are consistent). Finally, the \texttt{SSTORE} instruction (i.e., line 12 in Listing \ref{lst:B2X}) is executed to write the attacker's address to the slot where the owner is located, to obtain all permissions.

\subsection{Runtime Overhead}
DeFiTail mainly serves EVM-compliant blockchains, including Ethereum, Binance, EOS, Polygon, Arbitrum, Harmony, Fantom, and et al. While most blockchain platforms claim high transactions per second (TPS), it's crucial to recognize that most of these transactions involve interactions between user wallets rather than with smart contracts. 

\begin{table}[h]
% \vspace{-3ex}
\centering
\caption{The Comparison of Execution Speed of DeFiTail and Ethereum's TPS}
\label{tab:TPS_comparison}
\begin{tabular}{c | c | c | c}
    \hline
   \textbf{Situation} & \textbf{Tx Amount} & \textbf{ Time(s) } & \textbf{TPS} \\ 
    \hline
    $Ethereum_{all}$ & 436  &  30  &  14.53 \\
   $Ethereum_{sc}$  &  297  &  30  &  9.90  \\
   DeFiTail  &  321  &   19.27   &  $\approx$16.66\\ 
    \hline 
\end{tabular}
\end{table}

    For example, on April 23, 2025, Ethereum shows a reported TPS of 14.7~\cite{Etherscan}. As shown in Table \ref{tab:TPS_comparison}, we collect transaction data by randomly selecting a 30-second window, where the transaction amount is 436 and its TPS is 14.53 that are similar to 14.7. During the 30s period, we find that only 297 transactions interacted with smart contracts, with a more accurate TPS of 9.9 for contract interactions. 

    During the detection process of DeFiTail, we spent an average of 0.06004 seconds per transaction, with a $\approx$16.66 TPS, which is larger than the TPS in Ethereum.

\subsection{Threat to Validity}
Our work focuses on exploring malicious event detection on DeFi protocols.

\noindent\textbf{Internal validity: }In the process of gathering the dataset, we meticulously analyzed each security incident to extract the attack and victim contracts, ensuring the dataset's ground truth. Our dataset comprises 1,251 data paths for positive samples, which contains 692 data paths from flash loan attacks and 559 data paths from access control events. In addition, we randomly select 1,336 data paths for negative samples to ensure balance. The dataset is then divided into training and test sets in a 90\%: 10\% ratio.

\noindent\textbf{External validity: }Our dataset encompasses 25 blockchains, making it suitable for various blockchain environments. The contracts for all experiments are gathered from each EVM-compatible blockchain mainnet, and the security events are labeled by the REKT Database, rendering this data highly applicable for evaluation. 

\vspace{-1ex}

\subsection{Limitations}
During our evaluation, we lacked an evaluation of the impact and cost of false positives on the DeFiTail framework. Due to the high volume and frequency of transactions on Ethereum, we will involve huge interaction data in the detection process of DeFiTail, which makes it difficult for us to control the false positive rate, which is a disadvantage of DeFiTail. For example, during the 60 hours of monitoring in RQ4, DeFiTail examined about 1.12 million transactions. But to show DeFiTail's performance, we have tested the false negative rate in the testing set. In our results, we achieve a false positive rate of 1.616\% for access control and 2.585\% for flash loan exploitation.

Moreover, DeFiTail does not show the detailed details of the detected attacks, which makes users need to evaluate the vulnerability logic and its threat model. However, in the detection process, the interaction process involved in each exchange will be detected, which means that the scope of detection is in a transaction, and it is easier to determine whether there is relevant malicious behavior. We will optimize the output report in the future for better understanding.

DeFiTail uses historical data to learn potential exploitation patterns, making it limited to detecting emerging threats that are less frequent. However, through the evaluation of Section \ref{subsec::adaptability}, we find that DeFiTail can also find some new malicious exploitations in the real environment, which belong to access control and flash loan attacks. This illustrates that DeFiTail has the ability to detect new exploitations, which are subcategories of access control and flash loan exploitation.

\vspace{-1ex}

\subsection{Future Work}
Due to design principles and equipment limitations, DeFiTail only focused on the interaction relationship between two contracts on validated transactions. In the future, we will expand DeFiTail from the contracts and transactions aspects. In terms of contracts, we aim to enhance the chain-based interaction analysis capability of DeFiTail across more than three contracts. Regarding transactions, we will deploy full blockchain nodes to shorten the analysis time for pending transaction risks in the memory pool, ensuring that transactions are monitored before being packed into the blocks.

%% file: sections/relatedwork.tex
\section{Related Work}
\label{sec::relatedwork}
In this section, we primarily introduce the relevant work of the paper, which includes the studies on malicious DeFi behavior detection and smart contracts detection and analysis.

\subsection{Malicious DeFi Behavior Detection} 
The total amount of funds recovered in DeFi has risen to over \$77 billion~\cite{REKT-database2023solidity}, paralleling its rapid growth. During its evolution, DeFi protocols have been incorporated into several well-known blockchains. Smart contracts primarily control the management of users' funds, as these protocols span different blockchains. The interaction between different components creates notable vulnerabilities for potential attacks, which has had a significant impact on DeFi security research. Substantial research efforts~\cite{kong2023defitainter,zhang2023demystifying,tjiam2021your,wang2021blockeye,liu2022finding,ghaleb2023achecker,ramezany2023midnight,wang2021towards,qin2021attacking,li2024famulet,li2023demystifying,xie2024defort,liu2024experienced,yang2024hyperion,lin2024crpwarner,wu2024dappfl,su2023defiwarder,xu2023sok,xu2022reap, arora2024secplf} have been devoted to attacks on DApps or DeFi projects. 

First, there are sorts of studies that focus on the price manipulation or Oracle issues. Q, Kong, et al. \cite{kong2023defitainter} propose DeFiTainter, which uses tainted analysis techniques to detect price manipulation vulnerabilities in DeFi protocols. It traces cross-contracts by defining fixed interaction rules from call flow graphs and semantic induction. Z, Zhang, et al. \cite{zhang2023demystifying} conducted an empirical study examining 516 vulnerabilities in smart contracts between 2021 and 2022. It was found that 34.3\% of these vulnerabilities were related to price oracle manipulation. It was also concluded that auditing price oracle vulnerabilities would be more complex, as it would involve other types of real-world vulnerabilities. SecPlF~\cite{arora2024secplf} mainly protects Oracle from price manipulation attacks using flash lending, and realizes the protection against arbitrage attacks under the condition of low computational overhead. K, Tjiam, et al. \cite{tjiam2021your} analyze Oracle manipulation events in 2020-2021 using a lifecycle approach and summarize the effectiveness of potential countermeasures. B, Wang \cite{wang2021blockeye} proposes Blockeye, a two-step analysis process that combines symbol execution and transaction monitoring to examine Oracle contract state data and detect malicious transactions. J, Xu, et al~\cite{xu2023sok} study the state space model of decentralized exchanges, starting from AMM, analyzing the liquidity problem. J, Xu, et al~\cite{xu2022reap} analyze the algorithmic strategies of DeFi lending markets and liquidity pools, and finds that the volatility of token prices caused by complex contract patterns can cause severe losses.

Y, Liu, et al. \cite{liu2022finding} proposed SPCon, a solution for checking contracts for access control vulnerabilities. It uses role-mining technology to optimize problem-solving for permission issues. Achecker \cite{ghaleb2023achecker} uses static analysis of data flow and symbols to analyze permission problems from a contract perspective, resulting in improved performance. However, flash credit exploits~\cite{ramezany2023midnight,wang2021towards,qin2021attacking} on many crypto projects, such as Dapp or DeFi projects with complex functionality, have yet to be thoroughly researched. Midnight \cite{ramezany2023midnight} analyses the actions of events in transactions to detect the presence of flash credit exploits in specific transactions. D, Wang, et al. \cite{wang2021towards} and K, Qin, et al. \cite{qin2021attacking} both analyze the real flash loan attack patterns, optimize the efficiency of flash loan expoits, and detect the attack patterns.  DeFort~\cite{xie2024defort} framework targets price manipulation attacks in DeFi, which integrates price monitoring and profit calculation mechanisms to achieve attack tracking in different logics. Hyperion~\cite{yang2024hyperion} utilizes LLM to analyze the DApp front-end pages and identify the inconsistency between them and back-end data through symbolic execution techniques. Both CRPWarner~\cite{lin2024crpwarner} and Defiwarder~\cite{su2023defiwarder} constructed datalog detection rules for the Rug Pull problem in the DeFi domain, realizing automatic identification of malicious functions. 
The above methods bring advanced and excellent solutions or unique insights to various security problems in the DeFi field, and they all detect, locate, or defend possible vulnerabilities in the code. Different from them, DeFiTail tries to learn the interaction patterns between attacker and victim contracts in real malicious security events and monitor DeFi protocols.

\subsection{Smart Contract Analysis \& Vulnerability Detection} 
AI has made a significant impact in many industries, including detecting security vulnerabilities in the crypto projects. Various techniques (data process and AI) are being utilized to detect vulnerabilities in smart contracts \cite{chen2018detecting,zhuang2021smart,liu2021combining,gao2020deep,wu2021peculiar,sendner2023smarter,SmarTest2021UsenixSunbeom,he2019learning, wang2024smartinv, chen2024improving, chen2025chatgpt, chen2025forge}. Smart contracts can be processed through text sequences, tree structures, or graph structures and then fed into machine learning~\cite{chen2018detecting}, graph neural networks~\cite{zhuang2021smart,liu2021combining}, and inductive neural networks~\cite{gao2020deep,wu2021peculiar,sendner2023smarter} to detect vulnerable smart contracts automatically. W, Chen, et al. \cite{chen2018detecting} analyzed the transaction pattern of Ponzi vulnerabilities between accounts from transactions, then designed features and extracted these features into smart contracts for machine learning. The XGBoost algorithm detects Ponzi vulnerabilities in account and contract features. Z, Liu, et al. \cite{liu2021combining} embedded the smart contract syntax and integrated expert vulnerability detection capabilities into a graph neural network to detect vulnerabilities. ESCORT \cite{sendner2023smarter} first uses a specific encoder to extract the feature information in the contract. It then uses transfer learning with different models to classify the types of vulnerabilities from the features.
In addition, various techniques have been combined with AI to detect smart contracts, such as symbolic execution~\cite{SmarTest2021UsenixSunbeom} and fuzzing~\cite{he2019learning}. The ILF \cite{he2019learning} exploits the advantages of symbolic execution and fuzzing in deep learning. First, the transaction sequence is obtained through symbolic execution. It is then fed into the GRU model and combined with the fuzz mechanism to achieve improved inference and coverage. SmarTest \cite{SmarTest2021UsenixSunbeom} uses symbolic execution to obtain execution sequences and a well-defined language model to build a corpus that guides another symbolic execution to find vulnerable sequences. With the rise of large language models (LLMs), many studies explore the application of LLMs in the field of smart contract security~\cite{wang2024smartinv, chen2024improving, chen2025chatgpt, chen2025forge}. The Smartinv~\cite{wang2024smartinv} is a novel tool that uses multi-modal information to detect vulnerabilities by the Tier of Thought (ToT) prompt strategy. C. Chen, et al.~\cite{chen2025chatgpt} and J. Chen, et al.~\cite{chen2025forge} utilized LLMs to process smart contract vulnerabilities in batches, empirically exploring the role of LLMs in vulnerability detection.

In addition to employing AI methodologies, static and dynamic analysis techniques~\cite{ivanov2023txt} such as symbolic execution~\cite{luu2016making, bose2022sailfish,jin2022exgen,yang2023definition,zhang2024nyx}, and fuzz testing~\cite{liu2023rethinking,shou2023ityfuzz,chen2024towards,liang2025vulseye} are widely utilized for the identification of vulnerabilities in smart contracts. These methodologies provide robust frameworks for ensuring the security and integrity of blockchain-based applications. Different from them, DeFiTail uses heterogeneous graphs to capture features in long sequences in a short time, and fuses sequence execution features to obtain optimized contract execution logic features.

%% file: sections/conclusion.tex
\section{Conclusion}
\label{sec::conclusion}

The paper proposes DeFiTail, a novel framework for detecting DeFi protocols at the bytecode level. To capture the execution patterns of DeFi protocol attacks, we gather and connect the relevant contracts of the DeFi event attack process with the calling flows. To validate the correctness of CFG, we rely on the symbolic execution stack approach to verify the path feasibility of the connected CFG. To extract global features from extended sequences along the data path of the connected CFG, we construct a graph structure and utilize the BERT model to extract local sequential features. Subsequently, the combined features are used for classification detection. Experimental results reveal that our approach outperforms other SOTA tools in access control and flash loan exploits.